\documentclass[final,onecolumn,twoside]{IEEEtran}

\usepackage{amsmath, amsfonts, amssymb, mathrsfs}
\usepackage[amsmath,thmmarks]{ntheorem}
\usepackage{mathtools}
\usepackage{enumitem}
\usepackage{tikz}
\usepackage{color}
\usepackage{bbm}

\def\TT{\mathbb{T} }            	
\def\d{ \mathrm{d} }							
\def\I{ \mathrm{i} }							
\def\E{ \mathrm{e} }							
\def\NN{ \mathbb{N} }													
\def\ZN{ \mathbb{Z} }													
\def\RN{ \mathbb{R} }													
\def\QN{ \mathbb{Q} }													
\def\RNc{ \mathbb{R}_{\mathrm{c}} }						
\def\C{ \mathcal{C} }					  				
\def\CC{ \mathcal{C}_{\mathrm{c}} }			
\def\Cac{ \mathcal{C}_{\mathrm{ac}} }   
\def\W{ \mathcal{W} }										

\def\Mc{ \mathcal{M}_{\mathrm{c}} }
\def\Mcomp{ \mathcal{M}_{\mathrm{comp}} }
\def\Mnocomp{ \mathcal{M}_{\mathrm{nocomp}} }

\def\Uc{ \mathcal{U}_{\mathrm{c}} }
\def\Ud{ \mathcal{U_{\mathrm{d}}} }
\def\Ucomp{ \mathcal{U}_{\mathrm{comp}} }
\def\Unocomp{ \mathcal{U}_{\mathrm{nocomp}} }

\def\Vc{ \mathcal{V}_{\mathrm{c}} }
\def\Vcomp{ \mathcal{V}_{\mathrm{comp}} }
\def\Vnocomp{ \mathcal{V}_{\mathrm{nocomp}} }

\def\TM{ \mathrm{TM} }                  
\def\TMD{ \mathrm{TM_{\mathrm{D}}} }    

\def\bm{ \mathbf{m} } 

\newcommand{\Op}[1]{\mathrm{#1}}               
\DeclareMathOperator*{\suporinf}{\Theta}			 

\theoremstyle{plain}
\theoremheaderfont{\itshape\bfseries}
\theorembodyfont{\normalfont\itshape}
\theoremseparator{:}
\theoremsymbol{}
\newtheorem{theorem}{Theorem}[section]
\newtheorem{lemma}[theorem]{Lemma} 
\newtheorem{corollary}[theorem]{Corollary} 
\newtheorem{problem}[theorem]{Problem}
\newtheorem{defi}[theorem]{Definition}

\theoremstyle{plain}
\theoremheaderfont{\itshape}
\theorembodyfont{\normalfont}
\theoremseparator{:}
\theoremsymbol{}
\newtheorem{remark}{Remark}

\theoremheaderfont{\scshape\bfseries}
\newtheorem{example}{Example}

\begin{document}

\title{On non-detectability of non-computability and the degree of non-computability of solutions of circuit and wave equations on digital computers
\thanks{The work of H.~Boche was partly supported by the Deutsche Forschungsgemeinschaft (DFG, German Research Foundation) within the Gottfried Wilhelm Leibniz-program under Grant BO~1734/20--1,
under Germany's Excellence Strategy EXC-2092 CASA--390781972 and EXC-2111--390814868,
and by the German Federal Ministry of Education and Research (BMBF) within the national initiative on 6G Communication Systems through the research hub 6G-life under Grant 16KISK002, within the national initiative for Post Shannon Communication (NewCom) under Grant 16KIS1003K, and the project Hardware Platforms and Computing Models for Neuromorphic Computing (NeuroCM) under Grant 16ME0442.
V.~Pohl acknowledges the support of the DFG under Grant PO~1347/3--2.}
}

\author{Holger~Boche,~\IEEEmembership{Fellow, IEEE} and Volker~Pohl,~\IEEEmembership{Member, IEEE}
\thanks{
H. Boche is with the Chair of Theoretical Information Technology, Technical University of Munich, Arcisstra{\ss}e 21, 80333 M{\"u}nchen, Germany, e-mail: boche@tum.de.
He is also with the the BMBF Research Hub 6G-life,
with the Munich Center for Quantum Science and Technology (MCQST), 80799 M{\"u}nchen, Germany,
and with the Excellenzcluster Cyber Security in the Age of Large-Scale Adversaries (CASA), Ruhr Universit{\"a}t Bochum, Germany.
V.~Pohl is with the Chair of Theoretical Information Technology, Technical University of Munich, Arcisstra{\ss}e 21, 80333 M{\"u}nchen, Germany, e-mail: volker.pohl@tum.de.}
}

\date{21.04.2022}

\maketitle

\begin{abstract}
It is known that there exist mathematical problems of practical relevance which cannot be computed on a Turing machine.
An important example is the calculation of the first derivative of continuously differentiable functions.
This paper precisely classifies the non-computability of the first derivative, and of the maximum-norm of the first derivative in the Zheng-Weihrauch hierarchy.
Based on this classification, the paper investigates whether it is possible that a Turing machine detects this non-computability of the first derivative by observing the data of the problem,
and whether it is possible to detect upper bounds for the peak value of the first derivative of continuously differentiable functions.
So from a practical point of view, the question is whether it is possible to implement an exit-flag functionality for observing non-computability of the first derivative.
This paper even studies two different types of exit-flag functionality. A strong one, where the Turing machine always has to stop, and a weak one, where the Turing machine stops if and only if the input lies within the corresponding set of interest.

It will be shown that non-computability of the first derivative is not detectable by a Turing machine for two concrete examples, namely for the problem of computing the input--output behavior of simple analog circuits and
for solutions of the three-dimensional wave equation.
In addition, it is shown that it is even impossible to detect an upper bound for the maximum norm of the first derivative.
In particular, it is shown that all three problems are not even semidecidable.
Finally, we briefly discuss implications of these results for analog and quantum computing.
\end{abstract}


\section{Introduction}
\label{sec:Indro}

Computer simulations are essential tools in the design and development of systems for processing, transmitting, or storing information.
Because of the high complexity of modern information processing systems, and because of nonlinear components and algorithms, their behavior cannot usually be described by closed-form mathematical expressions.
Therefore, numerical simulations are extensively used for the design, development and analysis of such systems  \cite{Merhav_Simulation_TIT04,Fang_SimQuantumTIT20,Brambilla_Simulation_TCAS01,Chen_TCAD14}.
Although most physical models and systems are analog, i.e. operating with analog input and output quantities (e.g. time, voltage, current, etc.), simulations of such systems are usually done on digital computers, which are able to compute with finite sequences of rational numbers only.
So it is a natural and interesting question, which analog systems can actually be simulated on a digital machine \cite{Alur_DiscreteAbstraction2000,Cooper_ComputationalParadigms2008,BP1_ICASSP20}.
Moreover, investigations on the limits of present day digital computers is also of huge interest in view of the enormous progress made over the last years in the development of
new computing technologies, such as neuromorphic, analog, and quantum computing. 
These new computing approaches are theoretically as least as powerful as digital computers.
In fact, these new approaches have the potential to provide us with much more powerful hardware platforms, not only in the sense of complexity theory,
but in the sense that they will be capable of solving many more complex computational problems than present day digital hardware.

Groundbreaking works of Church \cite{Church_AJM1936}, Turing \cite{Turing_1937}, Specker \cite{Specker_1949} and many others showed that there exist fundamental limitations of digital computers with respect to their ability to calculate continuous (i.e. non-discrete) quantities.
Based on a theoretical model (the \emph{Turing machine}) for general digital computers,
several examples of non-computable operations were discovered in mathematics and computer science, e.g.:
\begin{itemize}
\item There are real numbers $r\in\RN$ which can't be computed on a Turing machine \cite{Specker_1949}.
\item There are computable continuous functions $u$ which are continuously differentiable but with a first derivative $u'(\tau)$ which is not computable on a Turing machine at certain computable points $\tau\in\RN$ \cite{Myhill_71}. 
\item There exist computable initial values for the three-dimensional wave equation so that the solution of the corresponding Cauchy problem is not computable at certain computable times and positions \cite{PourEl_WaveEqu81,PouEL_MathLogQuart97}.
\end{itemize}
Closely related to the mentioned results on the computability of the derivative is the computability theory of differential equations, and there already exist a quite advanced structure theory for the computability of different classes of differential equations, see, e.g.,  \cite{Garca_PolynomialDGL_2008} for polynomial differential equations and \cite{Garca_CompuIntervals_TAMS09} for general differential equations.
We also refer to the recent paper \cite{Graca_ComputabilityDiffEQU_2021} for an excellent review on this topic.

Apart from characterizing whether a number is computable or not, it is sometimes desirable to classify the non-computable numbers according to their degree of non-computability.
To this end, \emph{Zheng} and \emph{Weihrauch} introduced an arithmetic hierarchy of real numbers classifying different degrees of non-computability \cite{ZhengWeihrauch_MathLog01}.

From a practical side, recent years have seen a growing interest of computability problems in engineering and information theory. 
This research discovered more operations, important for practical applications, which are not computable even under reasonable physical assumptions regarding the data of the problem.
Examples include the calculation of channel capacities in information theory \cite{BSP_TrIT20},
the input-output relation of RLC circuits \cite{BP_IEEECAS20}, 
the calculation of Fourier- and Hilbert transforms \cite{PouEL_AdvMath83,BoMoTSP20_TuringBandlimited,BP_JAT2020_SpecFac},
the determination of spectral factorization \cite{BP_TransIT20},
and the calculation of the signal bandwidth \cite{BM_ComputabilityBandwidth21}.
Moreover there do exist standard approaches in information theory and signal processing
which naturally give a solution that is not effectively computable on digital hardware.
Examples include probabilistic methods \cite{Brown_IEEEIT_1972,BrownMorean_IEEEIT_1986},
optimization techniques \cite{CandesTao_IT05,Donoho_CS06,CandTao_IT10},
applications of distribution theory \cite{HeckelBoelcskei_TransIT13,PfanderWalnut_TRansIT16}, and 
sampling based techniques \cite{Beutler_CardinalSeries_TransIT76,Habib_RepresOperators_TIT01,PfanderWalnut_TRansIT06,BP_TransIT19}, to mention only a few.

\section{Motivation and Problem Formulation}
\label{sec:Motivation}

Non-computability has far reaching consequences for simulations with digital computers \cite{BP_IEEECAS20}.
If the output of an operation is not computable, then it is generally impossible to assess how well the simulated result approximates the real output.
This is of particular importance in computer-aided design (CAD) \cite{BauerLeclerc_TSP91,Gielen_ProcIEEE00,Goldgeisser_TCAS05}, where algorithms are expected to autonomously calculate certain quantities for given input data.
If the operation is not computable, these algorithms may fail to compute a solution that satisfy predefined error bounds.
To overcome these problems with non-computable solutions, one may try to algorithmically detect the non-computability before the actual computation.
Modern simulation programs often contain functionality to assess several aspects of the quality of the simulation. 
Such a quality assessment can be done before, during, or after the actual simulation. 
Then the simulation program may react accordingly.
For example: 
\begin{itemize}
\item Before the calculation, the program may check whether the input matrix for the simulation is ill-conditioned.
If so, the program may simply stop further calculations.
\item During the calculation, the program may observe whether the magnitude of the search direction becomes too small and stops further iterations.
\item During the calculation, the program may count iterations, and terminates if this number becomes too large.
\item After finishing the calculation, some algorithms can assess how confident the result might be.
\end{itemize}
These are just a few simple examples of the so-called \emph{exit-flag functionality} as can be found in software packages like \emph{MATLAB}, \emph{Mathematica}, \emph{Python}, and many others. 
With such a functionality, the simulation program itself is able to autonomously detect problems before/during/after the simulation, to make sure that the final result is sufficiently reliable.
However, from the point of view of computability, the above requirements on the exit-flag functionality are generally too strong, even through they reflect the actual approach taken in many practical applications, such as computer-aided system and signal design.
In these applications, one usually has the situation that either the program is able to solve the specific problem or, based on an exit flag functionality, it declares that it is not able to solve this specific problem.
The following example
shows that even very simple problems may not be algorithmically solvable but that it might be possible to slightly restrict the requirements to obtain a problem which is computationally solvable.

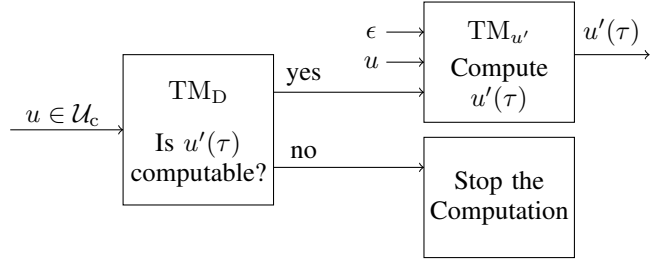
\begin{figure}
\begin{center}
\begin{tikzpicture}
	\draw[->] (-0.5,0) -- (1,0); \draw (0.2,0.2) node {$u\in\Uc$};
	\draw (1,-1) rectangle (3,1);
	\draw (2,0.5)  node {$\TM_{\mathrm{D}}$};
	\draw (2,-0.2)  node {Is $u'(\tau)$};
	\draw (2,-0.6)  node {computable?};
	\draw[->] (3.0,0.5) -- (5,0.5);     \draw (3.4,0.7)  node {yes};
	\draw[->] (4.5,0.9) -- (5,0.9);     \draw (4.3,0.9) node {$u$};
	\draw[->] (4.5,1.3) -- (5,1.3);     \draw (4.3,1.3) node {$\epsilon$};
	\draw[->] (3.0,-0.5) -- (5.0,-0.5); \draw (3.4,-0.3) node {no};
	\draw (5,0.1) rectangle (7,1.7);
	\draw (6,1.3)  node {$\TM_{u'}$};
	\draw (6,0.8)  node {Compute};
	\draw (6,0.4)  node {$u'(\tau)$};
	\draw (5.0,-1.7) rectangle (7.0,-0.1);
	\draw (6.0,-0.7) node {Stop the};
	\draw (6.0,-1.1) node {Computation};
	\draw[->] (7,1) -- (8,1);
	\draw (7.5,1.3) node {$u'(\tau)$};
\end{tikzpicture}
\end{center}
\caption{Exit-flag functionality for the differential operator: A Turing machine $\TM_{\mathrm{D}}(u)$ serves as a preprocessing unit for detecting whether $u'(\tau)$ is computable or not. 
Dependent on the result, it either stops the calculation or it starts a Turing machine $\TM_{u'}(u,\epsilon)$, which computes $u'(\tau)$ up to any predefined precision $\epsilon>0$.}
\label{fig:ProProc3}
\end{figure}

\begin{example}
\label{exa:ExampleNonCompu}
Assume that for a given computable number $x\in\RN$, we want to decide whether $x < 0$, whether $x=0$, or whether $x>0$. 
It is well known that this problem is not algorithmically solvable \cite{PourEl_Computability}, i.e. there exists \emph{no} Turing machine $\TM$ with input $x$ (or more precisely, with an algorithmic description of $x$) with $3$ possible output states ("$x<0$" , $x=0$, "$x>0$"), and which always stops for any arbitrary computable input $x\in\RN$.
The non-computability of this problem arises by the state "$x=0$".
Indeed, it was shown in \cite{PourEl_Computability} that there exists a Turing machine $\TM_{>0}$ with computable input $x \in \RN$ and which has only one output state ("$x>0$") and which stops if and only if $x>0$.
So the Turing machine $\TM_{>0}$ accepts exactly all computable numbers $x>0$.
In the same way, one shows that there exists a Turing machine $\TM_{<0}$ which accepts only all computable numbers $x<0$ \cite{PourEl_Computability}. 
Based on these two Turing machines one can now construct a third Turing machine $\TM_{*}$ whose inputs are all computational real numbers $x\neq 0$ and which will always stop in one of the two output states "$x<0$" or "$x>0$".
So by leaving out the state "$x=0$", we have obtained an algorithmically solvable problem.
\end{example}

\begin{figure}[t]
\begin{center}
\begin{tikzpicture}
	\draw[->] (-0.5,0) -- (1,0);   \draw (0.2,0.2) node {$u\in\Uc$};
	\draw (1,-1) rectangle (3,1);
	\draw (2,0.5)  node {$\TM_{\mathrm{P}}$};
	\draw (2,-0.2)  node {Is $u'(\tau)$};
	\draw (2,-0.6)  node {computable?};
	\draw[->] (3,0.5) -- (5,0.5);   \draw (3.4,0.7)  node {no};
	\draw[-] (3,-0.5) -- (3.8,-0.5);   \draw[->] (3.8,-0.5) -- (3.8,-1.2);   \draw (3.4,-0.3) node {yes};
	\fill[fill=blue!5] (2.2,-1.8) rectangle (5.4,-1.2);
	\draw (3.8,-1.5) node {$\TM_{\mathrm{P}}$ never halts.};
	\draw (5,-1) rectangle (7,1);
	\draw (6,0.3)  node {Stop the};	
	\draw (6,-0.3)  node {Calculation};
\end{tikzpicture}
\end{center}
\caption{Weak exit-flag functionality: The Turing machine $\TM_{\mathrm{P}}(u)$ serves as a preprocessing unit which halts only if $u'(\tau)$ is not computable for the given $u\in\Uc$. In this case, $\TM_{\mathrm{P}}$ stops the calculation with a notice that $u'(\tau)$ is not computable. Otherwise $\TM_{\mathrm{P}}$ runs forever.}
\label{fig:ProProc2}
\end{figure}
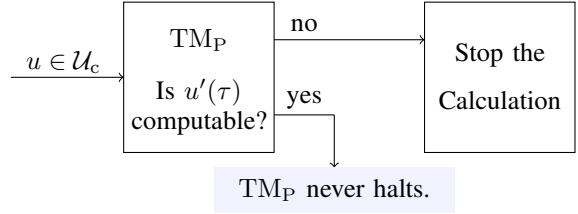

In this paper, we investigate problems similar to the following more concrete example: 
Assume a simulation program that has to determine the first derivative $u'(\tau)$ of a computable and continuously differentiable function $u$ at a computable point $\tau\in\RN$.
Does there exist an algorithm which is able to autonomously decide (without any human help), for any given $u$ in a certain subset $\Uc$ of computable and continuously differentiable functions, whether $u'(\tau)$ is computable or not? 
If such an algorithm exists, it could be used to implement an exit-flag functionality as sketched in Fig.~\ref{fig:ProProc3}.
Therein, a Turing machine $\TMD$ serves as a preprocessing unit which detects whether, for the given $u\in\Uc$, the value $u'(\tau)$ is computable.
If so, $\TMD$ starts the actual Turing machine $\TM_{u'}$ for calculating $u'(\tau)$ up to the given precision $\epsilon$.
Otherwise, if $u'(\tau)$ is not computable, $\TMD$ stops the calculation.
This paper is going to show that such a Turing machine $\TMD$ does not actually exist for a reasonable and practical relevant set $\Uc$.
%
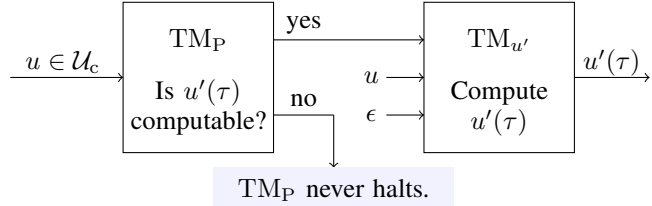
\begin{figure}
\begin{center}
\begin{tikzpicture}
	\draw[->] (-0.5,0) -- (1,0);   \draw (0.2,0.2) node {$u\in\Uc$};
	\draw (1,-1) rectangle (3,1);
	\draw (2,0.5)  node {$\TM_{\mathrm{P}}$};
	\draw (2,-0.2)  node {Is $u'(\tau)$};
	\draw (2,-0.6)  node {computable?};
	\draw[->] (3,0.5) -- (5,0.5);   \draw (3.4,0.7)  node {yes};
	\draw[-] (3,-0.5) -- (3.8,-0.5);   \draw[->] (3.8,-0.5) -- (3.8,-1.2);   \draw (3.4,-0.3) node {no};
	\fill[fill=blue!5] (2.2,-1.8) rectangle (5.4,-1.2);
	\draw (3.8,-1.5) node {$\TM_{\mathrm{P}}$ never halts.};
	\draw (4.3,0) node {$u$};
	\draw (4.3,-0.5) node {$\epsilon$};
	\draw[->] (4.5,0) -- (5,0);
	\draw[->] (4.5,-0.5) -- (5,-0.5);
	\draw (5,-1) rectangle (7,1);
	\draw (6,0.5)  node {$\TM_{u'}$};
	\draw (6,-0.2)  node {Compute};
	\draw (6,-0.6)  node {$u'(\tau)$};
	\draw[->] (7,0) -- (8,0);   \draw (7.5,0.2) node {$u'(\tau)$};
\end{tikzpicture}
\end{center}
\caption{Weak exit-flag functionality: A Turing machine $\TM_{\mathrm{P}}(u)$ serves as a preprocessing unit which halts only if $u'(\tau)$ is computable for the given $u\in\Uc$.
In this case, $\TM_{\mathrm{P}}$ starts the actual Turing machine $\TM_{u'}(u,\epsilon)$, which computes $u'(\tau)$ up to any predefined precision $\epsilon>0$.}
\label{fig:ProProc1}
\end{figure}
%
In fact, we will even show that the much weaker problem of detecting whether $u\in\Uc$ belongs to the set of functions for which $u'(\tau)$ is computable 
is not solvable by any Turing machine.
Such a Turing machine could be used for implementing a weak exit-flag functionality as sketched in Fig.~\ref{fig:ProProc1}.
Only if $u'(\tau)$ is computable, will a Turing machine be started for calculating $u'(\tau)$ up to the desired precision $\epsilon$.
Otherwise, if $u'(\tau)$ is not computable, the preprocessor $\TM_{\mathrm{P}}$ may never halt.
Alternatively, one may try to detect inputs $u\in\Uc$ for which the output $u'(\tau)$ is not computable. Then the Turing machine stops the calculation of $u'(\tau)$.
This alternative weak exit-flag functionality is sketched in Fig.~\ref{fig:ProProc2}.

\begin{remark}
Notice that the Turing machine $\TM_{*}$ in Example~\ref{exa:ExampleNonCompu} was obtained by combining the two Turing machines $\TM_{<0}$ and $\TM_{>0}$.
Each of these two machines implements, in fact, such a weak-exit flag functionality.
\end{remark}

\begin{example}
Another example where an exit flag functionality is actually not algorithmically computable comes from computer-aided circuit and system design.
Especially for non-linear circuits, the input-output behavior described by non-linear differential equations can often only be determined by simulations, because no closed-form solution exists.
In these situations, it is of central importance to determine the maximum (time) interval for which a solution exists.
If the input requires a solution outside this interval, the machine has to stop and has to indicate the this is not possible.
Nevertheless, as it was shown in \cite{Garca_CompuIntervals_TAMS09}, these intervals cannot algorithmically be determined and we refer to \cite{Graca_ComputabilityDiffEQU_2021} for an overview of more practically relevant results in this direction.
\end{example}

\subsection{Problem statement}

This paper investigates several aspects of the detectability of the non-computability of the first derivative of continuously differentiable functions.
The determination of the first derivative of certain signals or functions is an essential step in many algorithms in communications, signal processing, and quantum computing.
For example, to simulate the output of linear time-invariant systems described by an ordinary linear differential equation, one has to calculate the first (or higher order) derivative of the input signal.
In wireless communications, one often needs to dynamically determine the peak value of the transmit signal.
Applying the mean value theorem, this peak value can be estimated from the knowledge of the first derivative (see, e.g., \cite{WunderFischerBoche_OFDM2013} for an overview on this topic).
Moreover, new applications of quantum algorithms for breaking advanced cryptosystems also need the first derivative of a certain function as an input \cite{Eisentraeger_STOC14,BiasseSong_QuantumAlgorithm2016,BoerDucasFehr_QuantumComplex20}.
If this first derivative cannot be provided, these algorithms may fail to achieve the high computational efficiency compared to classical algorithms.

Since the first derivative of a computable continuously differentiable function is usually not computable,
it is natural to ask whether it is possible to algorithmically detect this non-computability. So we ask whether there exists an exit-flag functionality for detecting the non-computability of the first derivative.
This paper is going to show that this is actually not possible.
Nevertheless, in many applications (including the above examples from wireless communications and quantum computing), it will not be necessary to know the first derivative $u'(\tau)$ of a function $u$ at specific position $\tau\in\RN$ exactly, but it would be sufficient to have a good upper bound on the peak value of the first derivative, i.e. on $\left\|u'\right\|_{\infty}$.
The question then is, is it possible to algorithmically decide for a given computable and continuously differentiable function $u$ whether a certain number $\lambda$ is an upper bound for $\left\|u'\right\|_{\infty}$?
This paper is going to show that even this decision problem is generally not semidecidable.

\subsection{Contribution and outline of the paper}
\label{sec:Contribution}

In the first part of this paper, we investigate the computability of the first derivative for important subsets of continuously differentiable and computable functions. Starting with a known result on the non-computability of the first derivative, we will use the Zheng-Wheirauch hierarchy to precisely characterize the degree of this non-computability and the degree of the non-computability of the maximum norm of the first derivative. 
Since the first derivative and the maximum norm of the first derivative are generally not computable, the second part of this paper investigates whether it is possible to have exit-flag functionalities for detecting this non-computability.
In particular, we are going to show that it is impossible to detect:
\begin{itemize}
\item the non-computability of the first derivative,
\item an upper bound for the maximum norm of the first derivative,
\item the non-computability of the solution of the wave equation.
\end{itemize}
Thus, this paper answers two of Daniel Kahneman's questions \cite{Fridman_OnlineKahneman}, from the areas of semi-autonomous vehicles and general artificial intelligence.
Namely, he asks about problems which can be solved by humans but which can't be solved by machines, and he asks whether a machine is able to recognize by itself that it is faced with a problem which can only be solved by a human.

The remainder of the paper is organized as follows.
Section~\ref{sec:Notation} illustrates the importance of computability in circuit and system theory and it recalls 
the corresponding formal concepts and notions from computational analysis.
Section~\ref{sec:ComputeFirstDerivatice} precisely characterizes the non-computability of the first derivative, and of the maximum norm of the first derivative, for a subset of continuously differentiable and computable functions using the Zheng-Weihrauch hierarchy.
Section~\ref{sec:DecProblLTI} establishes the non-detectability of non-computability for the ordinary differential operator.
Section~\ref{sec:DetectionUpperBound} will show the impossibility of detecting an upper bound for the maximum norm of the first derivative, and
Section~\ref{sec:WaveEquation} establishes the non-detectability of the non-computability of solutions of the three-dimensional wave equation.
The paper closes with discussions of future problems in Section~\ref{sec:Conclusion}.
At several places in this paper, we will need properties of the Poisson integral.
These necessary results are collected in Appendix~\ref{sec:FirstAppendix}, at the end of the paper.

\section{Computability Theory}
\label{sec:Notation}

Computability theory plays a crucial role in assessing whether a continuous physical system can be simulated on a digital computer, and for the development of corresponding algorithms.
The first part of this section illustrates and explains this fundamental importance, based on a very simple but generic example from circuit theory.
Afterwards, we recall necessary notions and concepts from computability analysis to make the paper self-contained and accessible to readers not familiar with these concepts and notions.
Detailed introductions to the topic may be found in \cite{PourEl_Computability,HandbookCompTheory99,Weihrauch_ComputableAnalysis,AvigadBrattka_2014}.

\subsection{Circuit theory and computability}
\label{sec:CircuitsAndComputation}

\begin{figure}
\begin{center}
\begin{tikzpicture}
	\draw[thick] (0,1) -- (1.2,1);
	\draw[thick] (2.5,1) -- (5,1);
	\draw[thick] (0,-1) -- (5,-1);
	\draw[thick] (-0.1,1) circle (0.1);
	\draw[thick] (-0.1,-1) circle (0.1);
	\draw[thick] (5.1,1) circle (0.1);
	\draw[thick] (5.1,-1) circle (0.1);	
	\filldraw (3.8,1) circle (0.07);
	\filldraw (3.8,-1) circle (0.07);
	\draw[->,blue] (-0.1,0.8) -- (-0.1,-0.8);
	\draw[blue] (-0.7,0)  node {$u_{\mathrm{in}}(t)$};
	\draw[->,blue] (5.1,0.8) -- (5.1,-0.8);
	\draw[blue] (5.8,0)  node {$u_{\mathrm{out}}(t)$};
	\draw[thick] (3.8,-1) -- (3.8,-0.075);
	\draw[thick] (3.8,0.075) -- (3.8,1);
	\draw[very thick] (3.4,-0.075) -- (4.2,-0.075);
	\draw[very thick] (3.4,0.075) -- (4.2,0.075);
	\draw (3.3,-0.3)  node {\bfseries C};
	\draw[very thick] (1.2,0.75) rectangle (2.5,1.25);
	\draw (1.85,0.5)  node {\bfseries R};
\end{tikzpicture}
\end{center}
\caption{An RC--circuit composed of a resistor $R$ and a capacitor $C$ with input voltage $u_{\mathrm{in}}(t)$ and output voltage $u_{\mathrm{out}}(t)$.}
\label{fig:RCCircuit}
\end{figure}
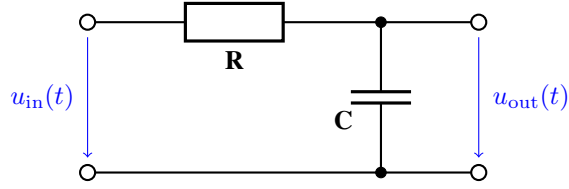

For motivation and illustration, we consider one of the main building blocks in circuit and system theory, namely a resistor--capacitor (RC) circuit as in Fig.~\ref{fig:RCCircuit}.
Its output voltage $u_{\mathrm{out}}$ is related to its input $u_{\mathrm{in}}$ by the first order differential equation
\begin{equation}
\label{equ:RC_DGL}
	RC\, \frac{\d u_{\mathrm{out}}}{\d t}(t) + u_{\mathrm{out}}(t) = u_{\mathrm{in}}(t)\,,
	\quad t \geq 0\,,
\end{equation}
with initial condition $u_{\mathrm{out}}(0) = u_{0}$.
Assume that we want to simulate this continuous system, i.e. for a given input $u_{\mathrm{in}}$, we want to calculate the value $u_{\mathrm{out}}(t)$ for times $t\geq 0$ exactly, on a digital computer.
For simplicity, we assume that $u_{\mathrm{in}}(t)$ is identical to zero and the initial condition is $u_{\mathrm{out}}(0) = u_{0} = 1$.
Then the solution of \eqref{equ:RC_DGL} is given by the closed-form expression
\begin{equation}
\label{equ:SolutionRC}
	u_{\mathrm{out}}(t) = \exp\left(-\tfrac{t}{RC}\right)\;,
	\quad t>0\,,
\end{equation}
and our goal is to calculate $u_{\mathrm{out}}(t)$ \emph{exactly} for $t > 0$.
Calculating $u_{\mathrm{out}}(t)$ exactly means that a digital computer is able to obtain the solution $u_{\mathrm{out}}(t)$ of \eqref{equ:RC_DGL} after finitely many computation steps.

We use the computing model of a \emph{Turing machine} \cite{Turing_1937}. 
It describes an upper limit for the capability of any digital computer.
If a problem is not solvable on a Turing machine, then it can't be solved on any digital computer.
A Turing machine can calculate exactly, but only with rational numbers. 
So all inputs have to be rational, and if the Turing machine has computed a certain number (after finitely many computation steps), then this number is again rational.
In our RC--circuit of Fig.~\ref{fig:RCCircuit}, we therefore assume that $R$ and $C$ are rational numbers,
and we restrict ourselves to calculate the output $u(t)$ at rational times $t>0$.
Then $t/(RC)$ is again rational and it follows from the Lindemann--Weierstrass theorem that the output $u_{\mathrm{out}}(t)$, given by \eqref{equ:SolutionRC}, is a transcendental number for every positive $t\in\QN$.
In other words, no Turing machine is able to simulate exactly the output of the RC--circuit in Fig.~\ref{fig:RCCircuit}.

Of course, from a practical point of view, there is no need to calculate $u_{\mathrm{out}}(t)$ exactly, but it is sufficient to obtain a (rational) approximation $\widetilde{u}_{\mathrm{out}}(t)$ of the exact (transcendental) solution $u_{\mathrm{out}}(t)$, but under an \emph{effective control} of the approximation error.
This means that we need a Turing machine $\TM$ which calculates in \emph{finite computation steps} a rational approximation $\widetilde{u}_{\mathrm{out}}(t)$, such that $\left| \widetilde{u}_{\mathrm{out}}(t) - u_{\mathrm{out}}(t) \right| < \epsilon$ with a predefined error bound $\epsilon\in\QN$, which we give to the Turing machine $\TM$.
If such a Turing machine exists, one would say that $u_{\mathrm{out}}(t)$ is \emph{(Turing) computable}.

For our simple example, the construction of such a Turing machine is fairly easy. 
It consists of an implementation of the power series expansion of the exponential function in \eqref{equ:SolutionRC} and uses the estimate of the remainder to effectively control the approximation error.
Nevertheless, for more complicated networks  or for more complicated input signals $u_{\mathrm{in}}$, it is generally no longer obvious whether a Turing machine for simulating this network can always be found.
In fact, it was shown in \cite{BP_IEEECAS20} that even for the simple RC-circuit in Fig.~\ref{fig:RCCircuit}, there exist computable, continuously differentiable input signals $u_{\mathrm{in}}$, so that the output $u_{\mathrm{out}}(t)$ is not computable for some computable time instances $t$.

So the requirement of an exact simulation of arbitrary analog circuits can be met only in very trivial cases (e.g. if the output signal is a polynomial). 
Even if the output can be expressed in "closed-form", i.e. in terms of, e.g., the exponential-, the sine-, or the cosine function, it can usually not be calculated exactly.
Therefore, one always needs to find techniques to approximate the exact solution under an effective control of the approximation error.
This paper deals with the question of whether it is possible to detect, algorithmically, that such an effective approximation is impossible.

\subsection{Notation and computable analysis}
\label{sec:CompAnal}

\paragraph*{General notation}
We primarily consider $2\pi$-periodic functions on $\RN$,
and so $\TT = \RN/2\pi\ZN$ often stands for the additive quotient group of real numbers modulo~$2\pi$.
Alternatively, $\TT$ sometimes denotes an arbitrary closed interval on $\RN$.
For any $1\leq p\leq\infty$, we write $L^{p}(\TT)$ for the usual Banach space of $p$-integrable functions on $\TT$ with norm $\left\| \cdot \right\|_{p}$, and
$\C(\TT)$ stands for the Banach space of continuous functions on $\TT$ with norm $\left\| f \right\|_{\infty} = \max_{t\in\TT} \left| f(t) \right|$.
The Banach space of all $k$-times continuously differentiable functions is denoted by $\C^{k}(\TT)$ and equipped with the norm 
$\left\| f \right\|_{\C^k} = \max_{n=0,1,\dots,k} \left\| f^{(n)} \right\|_{\infty}$.
We recall that a function $f$ on $\TT$ is said to be \emph{absolute continuous} if there exists a $g \in L^{1}(\TT)$, so that $f(t) = f(0) + \int^{t}_{0} g(\tau)\, \d\tau$ for all $t\in [0,2\pi)$.
Then the first derivative of $f$ exists almost everywhere on $\TT$, and $f' = g$ almost everywhere.
We write $\Cac(\TT)$ for the set of all absolute continuous functions on $\TT$.
The \emph{Fourier coefficients} of any $f \in L^{1}(\TT)$ are denoted by
$c_{n}(f) = \textstyle\frac{1}{2\pi} \int^{\pi}_{-\pi} f(\tau)\, \E^{-\I n \tau}\, \d\tau$, $n\in\NN$, and
the \emph{Wiener algebra} $\W$ is the set of all $f \in L^{1}(\TT)$ with an absolutely converging Fourier series, i.e.
\begin{equation*}
	\W = \left\{ f\in L^{1}(\TT)\ :\ \left\| f \right\|_{\W} = \textstyle\sum_{n\in\ZN} \left| c_{n}(f)\right| < \infty \right\}\;,
\end{equation*}	
which satisfies $\W \subsetneq \C(\TT)$ with $\left\| f \right\|_{\infty} \leq \left\| f \right\|_{\W}$ for all $f \in \W$.

\paragraph*{Computable analysis}
A sequence $\left\{ r_{n} \right\}_{n\in\NN}\subset\QN$ of rational numbers is said to be \emph{computable} if there exist recursive functions $a,b,s : \NN\to\NN$, such that $r_{n} = (-1)^{s(n)} a(n)/b(n)$
and with $b(n) \neq 0$ for all $n\in\NN$.
A \emph{recursive function} $a : \NN^{N} \to  \NN$ is a mapping that is built from elementary computable functions and recursion \cite{Soare_Recursively_Enumerable}.
For us, it is only important that a recursive function has an integer input, an integer output, and that it can be calculated on a \emph{Turing machine} \cite{Turing_1937,Turing_1938,Weihrauch_ComputableAnalysis}.
This is an abstract device which provides a theoretical model describing the fundamental limits of any realizable digital computer.

Any real number $x\in\RN$ is the limit of a sequence of rational numbers.
In order for $x\in\RN$ to be \emph{"computable"}, the convergence of the corresponding sequence of rationals has to be \emph{effective} in the following sense.
\begin{defi}[Effective convergence]
\label{def:effecConv}
A sequence $\left\{ x_{n} \right\}_{n\in\NN}$ in $\RN$ 
is said to \emph{effectively converge} to a limit $x \in \RN$ if there exists a recursive function $e : \NN\to\NN$ so that for all $N\in\NN$, one has
\begin{equation}
\label{equ:EffectiveConv}
	\left| x_{n} - x \right| \leq 2^{-N}
	\quad\text{for every}\ n \geq e(N)\,.
\end{equation}
\end{defi}
\begin{defi}[Computable numbers]
\label{def:CompNumber}
A number $x\in\RN$ is called \emph{computable} if there exists a computable sequence $\left\{ r_{n} \right\}_{n\in\NN} \subset\QN$ which effectively converges to $x$, and $\RNc$ will denote the set of computable real numbers.
\end{defi}

\begin{remark}
In Def.~\ref{def:effecConv} the approximation error \eqref{equ:EffectiveConv} is upper bounded by a function $\phi : \NN\to\RNc$ given by $\phi(N) = 2^{-N}$.
The concrete form of this function is of no importance for the definition of effective convergence.
Indeed $\phi(N)$ only needs to be monotonically decreasing in $N$ with $\lim_{N\to\infty} \phi(N) = 0$ and $\phi(N)$ needs to be a computable number for every $N\in\NN$.
\end{remark}

\begin{defi}[Computable functions]
\label{defi:ComputableFkt}
A function $f : \TT\to\RN$ is \emph{computable} if and only if there exists a computable sequence of real trigonometric polynomials $\left\{ p_{m} \right\}_{m\in\NN}$ which effectively converges to $f$ in the uniform norm, i.e. if there exists a recursive function $e : \NN\to\NN$, such that for all $t\in\TT$ and every $N\in\NN$,
\begin{equation*}
	m\geq e(N)
	\qquad\text{implies}\qquad
	\left| f(t) - p_{m}(t) \right| \leq 2^{-N}\;.
\end{equation*}
\end{defi}
So a function $f$ is computable if $f$ can effectively be approximated by trigonometric polynomials $p_{m}$ with computable coefficients.
Since $\left\{ p_{m} \right\}_{m\in\NN}$ converges uniformly, the limit $f$ is necessarily a continuous function.

\begin{defi}[Computability in a Banach space]
\label{defi:CompuBanachSpace}
A function $f$ in a Banach space $\mathcal{X}$ of functions on $\TT$ is said to be $\mathcal{X}$-computable if
there exists a computable sequence of real trigonometric polynomials $\left\{ p_{m} \right\}_{m\in\NN}$ 
and a recursive function $e : \NN\to\NN$, such that for every $N\in \NN$,
\begin{equation*}	
	\left\| f - p_{m} \right\|_{\mathcal{X}} \leq 2^{-N}
	\quad\text{for all}\ m\geq e(N)\;.
\end{equation*}
The set of all $\mathcal{X}$-computable functions will be denoted by $\mathcal{X}_{\mathrm{c}}$.
\end{defi}
The case $\mathcal{X} = \C(\TT)$ is of particular interest to us.
It follows from the uniform and effective convergence in Def.~\ref{defi:ComputableFkt} that every computable function is also $\CC(\TT)$-computable.

In the following, we consider Turing machines $\TM$ whose inputs $x$ are either computable numbers or computable functions.
It should be noted that these inputs are always given to the Turing machine in the form of a program, which allows $\TM$ to compute $x$ effectively.
For example, if we want to give the computable number $x = \pi$ to a Turing machine $\TM$, then we hand over a description of $\pi$, i.e. a program which can be run on $\TM$ and which effectively computes a rational approximation of $\pi$ up to any necessary precision.

\begin{remark}
This form of representing all the information about $x$ by an algorithm, which is able to compute for every arbitrary given error a corresponding rational approximation of $x$, is most appropriate for digital computing hardware.
We only mentioned that there exist other computing concepts which allow, in principle, the computation with real numbers (e.g. Blum-Shub-Smale machines \cite{BSSComputingModel_89}). However, these concepts are not relevant for digital hardware.
These computing concepts differ in the way information about numbers is represented.
\end{remark}

\subsection{Decidability and deciders}
\label{sec:Decidability}

Subsequently, it will be important to distinguish between \emph{decidable} and \emph{semidecidable} sets and to realize that every such set is associated with a particular type of Turing machine.
To this end, we first recall that a subset $A \subset \NN$ of natural numbers is said to be \emph{decidable} (or \emph{recursive}) if there exists an algorithm taking natural numbers as input and which correctly decides, after finitely many iterations, whether the input belongs to $A$ or to the complement $A^{\mathrm{c}} = \NN\backslash A$.
A set $A \subset \NN$ is called \emph{semidecidable} (or \emph{recursively enumerable}) if there is an algorithm which correctly identifies (after finitely many iterations) all inputs belonging to $A$, but which may not halt if the input does not belong to~$A$.
Since the set $\RNc$ of all computable numbers is countable, the previous definition of decidable and semidecidable sets also applies to subsets of $\RNc$.
However, we will later consider subsets $B$ which are known to belong to a certain subset $\mathcal{U}$ of $\RNc$.
Knowing that $B$ belongs to $\mathcal{U} \subset\RNc$ may make it simpler to decide whether a given $x \in \mathcal{U}$ belongs to $B$ or not.
So we will need the following notion of decidablility relative to a basic set $\mathcal{U}$.

\begin{defi}[Decidable and semidecidable sets]
\label{defi:DecidableSets}
Let $\mathcal{U} \subset \RNc$ be a subset of computable numbers. 
A set $B\subset\mathcal{U}$ is said to be \emph{decidable} (with respect to $\mathcal{U}$) if there exists a Turing machine $\TM(x)$ with input $x\in\mathcal{U}$ which correctly identifies, after finitely many iterations, whether $x \in B$ or $x \in B^{\mathrm{c}} = \mathcal{U}\backslash B$.

A set $B\subset\mathcal{U}$ is said to be \emph{semidecidable} (with respect to $\mathcal{U}$) if there exists a Turing machine $\TM(x)$ with input $x \in \mathcal{U}$
which correctly identifies (after finitely many iterations) every input $x \in B$, but which may run forever for $x \in B^{\mathrm{c}} = \mathcal{U}\backslash B$.
\end{defi}

\begin{remark}
We again emphasize the importance of the basic set $\mathcal{U}$ in Def.~\ref{defi:DecidableSets}.
If, for example, the basic set is reduced to a subset $\mathcal{U}_{1}\subset\mathcal{U}$, then this means that one has more information about $B\subset \mathcal{U}_1$.
Consequently, a subset $B \subset \mathcal{U}_{1}$ which is not semidecidable with respect to $\mathcal{U}$ may be semidecidable with respect to $\mathcal{U}_{1}$.
In most cases the set $\mathcal{U}$ will be clear from the context, and so we can often omit the addition "with respect to $\mathcal{U}$".
\end{remark}

For the semidecidability of $B$, Def.~\ref{defi:DecidableSets} requires only very weak properties on the corresponding Turing machine $\TM(x)$.
This Turing machine has only \emph{one output state} and it is guaranteed to halt only if the input $x\in\mathcal{U}$ belongs to the subset $B\subset\mathcal{U}$.
Otherwise, if $x\notin B$, $\TM$ might never halt.
Clearly, it is not possible to observe that $\TM$ runs forever. 
One can only notice that $\TM(x)$ still runs after $N$ iterations. 
But this tells us nothing about $x$, because $\TM(x)$ might halt after $N+k$ iterations, or it might run forever.
Only if $\TM(x)$ terminates its calculation, can we be sure that $x$ belongs to $B$. 
The situation is completely different for a decidable set $B$.
In this case, $\TM(x)$ always halts (after a finite number of iterations) and it has \emph{two output states}, namely "$1$" if $x$ belongs to the subset $B\subset \mathcal{U}$ and "$0$" if $x$ belongs to the complement $B^{\mathrm{c}} = \mathcal{U}\backslash B$.
So in this case one only has to wait until $\TM(x)$ halts.
Then its output shows whether $x$ belongs to $B$ or $B^{\mathrm{c}}$.
Turing machines corresponding to decidable sets are often called \emph{deciders}, and if we want to emphasize that a Turing machine is a decider, we will denote it by $\TMD$.

\subsection{Degrees of non-computability: Zheng-Weihrauch hierarchy}
\label{sec:NotationWZ}

A real number $x\in\RN$ is computable (cf. Def.~\ref{def:CompNumber}) if there exists a computable sequence $\left\{ r_{n} \right\}_{n\in\ZN}$ of rational numbers which effectively converges to $x$.
Here the \emph{effectivity of convergence} (cf. Def.~\ref{def:effecConv}) is important, since there exist computable sequences which do not converge effectively to an $x\in \RN$.
In this case, $x$ is \emph{non-computable}.
The standard example is a so-called \emph{Specker sequence} \cite{Specker_1949}: If $A \subset \NN$ is a non-recursive set, then the real number $x = x[A] = \sum_{n\in A} 2^{-n}$ is known to be non-computable.
 
In many situations, it is not only important to know that a number is non-computable, but how much worse this non-computability actually is.
In order to characterize such different degrees of non-computability, Zheng and Weihrauch introduced an arithmetical hierarchy of real numbers \cite{ZhengWeihrauch_MathLog01}.
In this hierarchy, we only need the first two classes.
Therefore, following \cite{ZhengWeihrauch_MathLog01}, we briefly give an intuitive explanation of these two classes, before giving the full definition of the \emph{Zheng-Weihrauch hierachy} for the sake of completeness.
We denote by $\Sigma_{1}$ ($\Pi_{1}$), the set of all $x \in \RN$, which is the limit of an increasing (decreasing) computable sequence $\left\{ r_{n} \right\}_{n\in\ZN} \subset \QN$, and  $\Delta_{1} = \Sigma_{1} \cap \Pi_{1}$. 
The question is now whether all elements of $\Sigma_{1}$ (or $\Pi_{1}$) are computable. 
The answer is no, and it is actually known that $\Delta_{1}$ is precisely the set of all computable numbers.
Note that the previous example of the Specker sequence shows that generally, $\Sigma_{1}\backslash\Delta_{1}  \neq \emptyset$, because $x[A]$ is the limit of a monotonically increasing sequence.
Similarly, one also has $\Pi_{1}\backslash \Delta_{0} \neq \emptyset$.
Now, one may generalize the above question by considering the sets $\Sigma_{2}$ and $\Pi_{2}$ of all real numbers $x = \limsup_{n\to\infty} r_{n}$ and $x = \liminf_{n\to\infty} r_{n}$, respectively, with a computable sequence of rational numbers $\left\{ r_{n} \right\}_{n\in\NN}$.
One may ask whether every $x \in \Delta_{2} = \Sigma_{2} \cap \Pi_{2}$ is computable.
Here the answer is no, i.e. there exist numbers $x\in\Delta_{2}$ so that $x\notin\Sigma_{1}\cap\Pi_{1}$.

Continuing the above process, \emph{Zheng} and \emph{Weihrauch} introduced the following hierarchy of sets $\Sigma_{n}$, $\Pi_{n}$ and $\Delta_{n}$:

\begin{defi}[Zheng-Weihrauch Hierachy]
For $n\in\NN$ let $\bm = (m_1, \dots, m_{n}) \in \NN^n$.
Then the sets $\Sigma_{n} \subsetneq \RN$, $\Pi_{n} \subsetneq \RN$, and $\Delta_{n} \subsetneq \RN$ are defined as follows:
\begin{itemize}
\item A number $x \in \RN$ belongs to $\Sigma_{n}$ if and only if there exists an $n$-fold computable sequence $\left\{ r_{\bm} \right\}_{\bm\in\NN^{n}}$ of rational numbers such that
\begin{equation*}
	x = \sup_{m_{1}\in\NN} \inf_{m_{2}\in\NN} \dots \suporinf_{m_{n}\in\NN} r_{\bm}\;,
\end{equation*}
and where $\suporinf$ stands for the infimum if $n$ is even and for the supremum if $n$ is odd.
\item A number $x \in \RN$ belongs to $\Pi_{n}$ if and only if there exists an $n$-fold computable sequence $\left\{ r_{\bm} \right\}_{\bm\in\NN^{n}}$ of rational numbers, such that
\begin{equation*}
	x = \inf_{m_{1}\in\NN} \sup_{m_{2}\in\NN} \dots \suporinf_{m_{n}\in\NN} r_{\bm}\;,
\end{equation*}
and where $\suporinf$ stands for the supremum if $n$ is even and for the infimum if $n$ is odd.
\item $\Delta_{n} = \Sigma_{n} \cap \Pi_{n}$.
\end{itemize}
\end{defi}

\begin{remark}
We note that for every $n\geq 1$, one has $\Delta_{n} \subsetneq \Sigma_{n}$ and $\Delta_{n} \subsetneq \Pi_{n}$ and $\Delta_{n} \subsetneq \Delta_{n+1}$. Since $\Delta_{1} = \RNc$,
the sets $\Delta_{n}$ describe subsets of $\RN$ whose entries are increasingly less computable as $n$ increases.
\end{remark}

\begin{remark}
There is a very close relation between the arithmetical hierarchy of Zheng and Weihrauch and the \emph{Kleene-Mostowski hierarchy} of subsets of natural numbers \cite{Kleene_TRansAMS43,Mostowski_FundMath47}.
\end{remark}
According to \cite{ZhengWeihrauch_MathLog01}, to every $x \in \Sigma_{1} \cap [0,1]$ there exists a unique recursively enumerable set $A\subset \NN$ so that
\begin{equation}
\label{equ:xA}
	x = x[A] = \textstyle\sum_{n\in A} 2^{-n}\;.
\end{equation}
Conversely, every recursively enumerable set $A\subset \NN$ defines by \eqref{equ:xA} a number $x \in \Sigma_{1} \cap [0,1]$.
Since \eqref{equ:xA} converges absolutely, it can be rearranged, which gives rise to the following definition.
\begin{defi}[Zheng-Weihrauch Representation]
Let $x\in \Sigma_{1} \cap [0,1]$ be arbitrary with a corresponding recursively enumerable set $A\subset \NN$, such that \eqref{equ:xA} holds.
Then every one-to-one mapping $\varphi_{A} : \NN \to A$ is said to be a \emph{Zheng-Weihrauch representation} of $x$.
\end{defi}
In other words, $\varphi_{A} : \NN \to A$ is a Zheng-Weihrauch representation of $x$ if and only if $x = \sum_{n\in\NN} 2^{-\varphi_{A}(n)}$.

\section{Computability of the First Derivative}
\label{sec:ComputeFirstDerivatice}

The differential operator $\Op{D} : u \mapsto u'$ is a fundamental operator in circuit and system theory because the voltage and current on an ideal capacitor or inductor are related by 
$i(t) = C\, u'(t)$ or $u(t) = L\, i'(t)$, respectively.
Simulating larger networks or circuits usually yields a higher order ordinary differential equation. 
To keep the derivations as simple as possible, here we consider only the first derivative, but it will be clear that the results extend to higher order differential equations.
In the first subsection, we recall that for an important class of continuously differentiable functions, the first derivative is generally non-computable.
In the second subsection, we then show that $\left\|u'\right\|_{\infty}$ is also not computable and we 
characterize the degree of the non-computability of $u'$ and $\left\|u'\right\|_{\infty}$ in the Zheng-Weihrauch hierarchy.

\subsection{Degree of non-computability of $u'$}

This section considers functions on $\TT = \RN/2\pi\ZN$, and 
it is known that there exist computable continuous functions which are continuously differentiable ($u\in\CC(\TT)\cap\C^{1}(\TT)$), but which have a first derivative $u'$ which is not a computable function \cite{Myhill_71}.
Even more, there are functions $u\in\CC(\TT)\cap\C^{1}(\TT)$ satisfying additional smoothness conditions, but such that $u'(\tau)$ is still not computable for some $\tau\in\TT$.
Since this result from \cite{BP_IEEECAS20} will be the basis for the following investigations, we briefly restate it for future reference.
To this end, we define the set
\begin{equation}
\label{equ:Def_Uc}
	\Uc = \left\{ u\in\CC(\TT)\cap\C^{1}(\TT)\ :\ u'\in\Cac(\TT) \cap \W \right\}
\end{equation}
of all computable continuous functions on $\TT$ with an absolute continuous first derivative which belongs to the Wiener algebra $\W$.
Then \cite{BP_IEEECAS20} gives an explicit construction of a $u_{A}\in\Uc$, such that $u'_{A}(0)$ is not a computable number.

\begin{theorem}
\label{thm:MainNegativResult}
Let $A \subset\NN$ be a recursively enumerable set which is not recursive and define
\begin{equation}
\label{equ:function_uA}
	u_{A}(t)
	= \sum^{\infty}_{n=1} 2^{-\varphi_{A}(n)}\, p_{n}(t)\,,
	\quad t\in\TT\,,
\end{equation}
with trigonometric polynomials $p_{n}$ given for $n\in\NN$ by
\begin{equation}
\label{equ:trigPoly_pn}
	p_{n}(t)
	= \frac{1}{G(n)} \sum^{n+1}_{k=2} \frac{\sin(kt)}{k^{2}\log k}\;,
	\quad t\in\RN\;,
\end{equation}
and with computable numbers $G(n) = \textstyle\sum^{n+1}_{k=2}\frac{1}{k\log k}$.
Then $u_{A} \in \Uc$, but $u'_{A}(0) = \left\| u'_{A} \right\|_{\infty}\notin\RNc$.
\end{theorem}

\begin{remark}
It might be worth noting that the function $u_{A}$ constructed in Theorem~\ref{thm:MainNegativResult} has the property that
$u'_{A}(t) \in \RNc$ for every $t \in \TT \cap \RNc$, $t\neq 0$ but $u'_{A}(0)\notin\RNc$ \cite{BP_IEEECAS20}.
\end{remark}
It is clear that Theorem~\ref{thm:MainNegativResult} might be reformulated for arbitrary points on $\TT$:
\begin{corollary}
For any $\tau\in\TT\cap\RNc$ there exists a $u_{A,\tau}\in\Uc$, so that $u'_{A,\tau}(\tau) = \left\| u'_{A,\tau} \right\|_{\infty} \notin\RNc$.
\end{corollary}
Without loss of generality and to simplify notation, we stick to the case $\tau=0$ as given in Theorem~\ref{thm:MainNegativResult}.

\begin{remark}
We emphasize that $u_{A}$, defined in \eqref{equ:function_uA}, is a computable continuous function. 
Therefore, in view of Lemma~\ref{defi:ComputableFkt}, there exists a sequence $\left\{ u_{m} \right\}_{m\in\NN}$ of trigonometric polynomials, which converges uniformly and effectively to $u_{A}$. 
In particular, it was shown in \cite{BP_IEEECAS20} that
\begin{equation}
\label{equ:uA_um}
	\left| u_{A}(t) - u_{m}(t) \right|
	< \frac{K_{0}}{G(m+1)}\,,
	\quad\text{for every}\ t\in\TT\,,
\end{equation}
with a computable positive constant $K_{0}\in\RNc$ and with trigonometric polynomials
\begin{equation}
\label{equ:um}
	u_{m}(t) = \textstyle\sum^{m}_{n=1} 2^{-\varphi_{A}(n)}\, p_{n}(t)\,,
	\quad t\in \TT\,,
\end{equation}
which are the partial sums of \eqref{equ:function_uA}.
\end{remark}

\subsection{Degree of computability of $u'$ and $\left\| u' \right\|_{\infty}$}
\label{sec:InftyNorm}

Theorem~\ref{thm:MainNegativResult} showed that there exist functions $u \in \Uc$ so that $u'(0)$ and $\left\|u'\right\|_{\infty}$ are not computable numbers.
Next, we want to determine the degree of non-computability of these two numbers using the Zheng-Weihrauch hierarchy.
Later, in Section~\ref{sec:DetectionUpperBound} we use these results to show that the problem of algorithmically detecting an upper bound for $\left\|u'\right\|_{\infty}$ is not semidecidable, 
and we will discuss consequences for recent quantum algorithms.

\begin{lemma}
\label{lem:DegreeComUstrich}
Let $u\in\Uc$ be arbitrary, then $u'(t) \in \Delta_{2}$ for every $t\in\RNc$.
\end{lemma}

\begin{IEEEproof}
The function $u$ is continuously differentiable and $2\pi$-periodic.
So for every fixed $t\in\RNc$, one can define a computable sequence $\left\{ r_{n} \right\}_{n\in\NN}$ of computable numbers by
\begin{equation*}
	r_{n}(t) = \frac{u(t + 1/n) - u(t)}{1/n}\;,
	\quad n\in\NN\;,
\end{equation*}
which satisfies $\lim_{n\to\infty} r_{n}(t) = u'(t)$.
Since for every convergent sequence its limit inferior and and limit superior are equal, we have $u'(t) \in \Delta_{2}$.
\end{IEEEproof}

\begin{remark}
Together with Theorem~\ref{thm:MainNegativResult}, Lemma~\ref{lem:DegreeComUstrich} shows in particular that not all numbers $x \in \Delta_{2}$ are necessarily computable.
\end{remark}

Next we characterize the degree of non-computability for the maximum norm of the first derivative for functions in $\Uc$.
The following theorem shows that for every $u\in \Uc$, the maximum norm $\left\|u'\right\|_{\infty}$ is the limit of a monotone increasing computable sequence of computable numbers, so 
$\left\|u'\right\|_{\infty} \in \Sigma_{1}$.
\begin{theorem}
\label{thm:InfNormSigma1}
Let $u \in \Uc$ be arbitrary, then $\left\| u' \right\|_{\infty} \in \Sigma_{1}$.
\end{theorem}

\begin{IEEEproof}
Let $u\in\Uc$ be arbitrary. Then both, $u$ and $u'$ are continuous functions on $\TT$.
Consequently, $u$ can be written as a \emph{Fourier series}
\begin{equation*}
	u(t) = \textstyle\frac{a_{0}}{2} + \sum^{\infty}_{n=1} \left[ a_{n} \cos(nt) + b_{n} \sin(nt) \right]\;,
	\quad t\in\TT\;,
\end{equation*}
which converges uniformly on $\TT$ and with the \emph{Fourier coefficients}
\begin{equation*}
	a_{n} = \frac{1}{\pi}\int_{\TT} u(\tau) \cos(n\tau)\, \d\tau
	\ \text{and}\ 
	b_{n} = \frac{1}{\pi}\int_{\TT} u(\tau) \sin(n\tau)\, \d\tau .
\end{equation*}
Both $\left\{ a_{n} \right\}_{n\in\NN}$ and $\left\{ b_{n} \right\}_{n\in\NN}$ are computable sequences of computable numbers.
Then for $0\leq r < 1$ the Poisson integral \eqref{equ:PoissonIntegral} of $u$ is given by
\begin{equation}
\label{equ:PoissonSum}
	\left( \Op{P}_r u\right)(t)
	= \textstyle\frac{a_{0}}{2} + \sum^{\infty}_{n=1} r^n \left[ a_{n} \cos(nt) + b_{n} \sin(nt) \right]
\end{equation}
for $t\in\TT$.
Since $u$ is a computable continuous function, it follows that for every $r \in [0,1] \cap \RNc$, the mapping $u \mapsto \Op{P}_{r} u$ from $\C(\TT)$ to $\C(\TT)$ is computable.
Moreover, since $u'$ is continuous on $\TT$, one has for every $r \in [0,1]$
\begin{equation*}
	\left( \Op{P}_{r} u' \right)(t) = \frac{\d}{\d t} \left(\Op{P}_{r} u\right)(t)\;,
\end{equation*}
which easily follows from the representation \eqref{equ:PoissonSum} of the Poisson integral $\Op{P}_{r} u$.
Moreover, for every $r \in [0,1] \cap \RNc$, the mapping $\frac{\d}{\d t} \circ \Op{P}_{r} : \C(\TT) \to \C(\TT)$ is also computable \cite{PourEl_Computability}.
Therefore with $r_{n} = 1 - 1/n$,
\begin{equation*}
	d_{n}(f)
	= \max_{t\in\TT} \left| \left(\Op{P}_{r_{n}} u'\right)(t) \right|
	= \max_{t\in\TT} \left| \tfrac{\d}{\d t}\left(\Op{P}_{r_{n}} u\right)(t) \right|
\end{equation*}
defines a computable sequence $\left\{ d_{n} \right\}_{n\in\NN}$ of computable numbers.
Moreover, it follows from \eqref{equ:PoissonSum} that for all $r,\varrho \in [0,1]$, one has $\Op{P}_{\varrho \cdot r} u = \Op{P}_{\varrho}\left(\Op{P}_{r} u\right)$.
So for arbitrary $0 \leq r_{1} < r_{2} \leq 1$ and with $\mu = r_{1}/r_{2}$, one gets
\begin{align*}
	\max_{t\in\TT} \left|\left(\Op{P}_{r_1} u\right)(t)\right|
	&= \max_{t\in\TT} \left|\left(\Op{P}_{\mu r_2} u\right)(t)\right|
	= \max_{t\in\TT} \left|\left(\Op{P}_{\mu}\left(\Op{P}_{r_2} u\right)\right)(t)\right|\\
	&\leq \max_{t\in\TT} \left|\left(\Op{P}_{r_2} u\right)(t)\right|\;,
\end{align*}
where the last inequality follows from the maximum-modulus principle \eqref{equ:maxModulusPoission}.
So because $r_{n} < r_{n+1}$ for all $n\in\NN$, one has $d_{n}(f) \leq d_{n+1}(f)$ for all $n\in\NN$, i.e. $\left\{ d_{n} \right\}_{n\in\NN}$ is a non-decreasing sequence.
In addition, $\lim_{n\to\infty} d_{n}(f) = \left\| u' \right\|_{\infty}$.
Thus $\left\| u' \right\|_{\infty}$ is the limit of a non-decreasing computable sequence of computable numbers, i.e. $\left\| u' \right\|_{\infty} \in \Sigma_{1}$.
\end{IEEEproof}

So for every $u\in \Uc$, one always has that $\left\|u'\right\|_{\infty} \in \Sigma_{1}$.
It is worth noting that the converse of this statement is also true:
Given an arbitrary positive number $x\in\Sigma_{1}$, there always exists a function $u\in\Uc$ so that $\left\| u'\right\|_{\infty} = x$. This is shown by the following theorem.

\begin{theorem}
\label{thm:InfNorm_FktUc}
Let $x \in \Sigma_{1}$, $x\geq 0$ be arbitrary, then there exists a $u \in \Uc$ with $\left\| u' \right\|_{\infty} = x$.
\end{theorem}

\begin{IEEEproof}
If $x=0$, then the function $u(t) = 0$ for all $t\in\TT$ has the properties required by the theorem.
Assume now $x \in \Sigma_{1}$, $x>0$.
We are going to construct $u\in\Uc$ with $\left\| u' \right\|_{\infty} = x$.
To this end, we first note that since $x \in \Sigma_{1}$, there exists a computable sequence $\left\{ r_{n} \right\}_{n\in\NN}$ of rational numbers with $r_{n} \leq r_{n+1}$ for all $n\geq 1$ and with $\lim_{n\to\infty}r_{n} = x$.
Without loss of generality, we assume $r_{0} = 0$ and define $d_{n} = r_{n} - r_{n-1}$ for all $n>1$.
Then $d_{n} \geq 0$ for all $n\in\NN$ and $\sum_{n\in\NN} d_{n} = \lim_{n\to\infty} r_{n} = x$.
Now we choose a computable sequence $\left\{ n_k \right\}_{k\in\NN} \subset \NN$ so that $G(n_{k}) > k^{2}$, wherein $G(n) = \sum^{n+1}_{k=2} \frac{1}{k\log k}$ as in the definition of the polynomials $p_{n}$ in \eqref{equ:trigPoly_pn}.
Therewith, we define the function
\begin{equation*}
	u(t) = \textstyle\sum^{\infty}_{k=1} d_{k}\, p_{n_k}(t)\,, 
	\quad t\in\TT\;.
\end{equation*}
The corresponding sequence $\left\{ P_{K} \right\}_{K\in\NN}$, given by $P_{K}(t) = \textstyle\sum^{K}_{k=1} d_{k}\, p_{n_k}(t)$, is a computable sequence of polynomials with
\begin{equation*}
	\left\| u - P_{K} \right\|_{\infty}
	\leq \textstyle\sum^{\infty}_{k=K+1} d_{k}\, \left\|p_{n_k}\right\|_{\infty}
	\leq C_{1} \textstyle\sum^{\infty}_{k=K+1} \frac{d_{k}}{G(n_{k})}\;,
\end{equation*}
using $\left| p_{n}(t) \right| \leq C_{1}/G(n)$ for all $t\in\TT$ and with the constant $C_{1} = \sum^{\infty}_{k=2}1/(k^2\log k) < \infty$.
Moreover, there always exists an $n_{*} \in \NN$ so that $n_{*} > x$.
Therefore $n_{*} > d_{k}$ for all $k\in\NN$, and so for every $K\in\NN$, we have
\begin{align*}
	\left\| u - P_{K} \right\|_{\infty}
	&\leq C_{1} n_{*} \textstyle\sum^{\infty}_{k=K+1} \frac{1}{G(n_{k})}\\
	&\leq C_{1} n_{*} \textstyle\sum^{\infty}_{k=K+1} \frac{1}{k^{2}}
	\leq C_{1} n_{*} \textstyle\int^{\infty}_{K} t^{-2} \d t
	\leq \frac{C_{2}}{K}\;,	
\end{align*}
showing that $P_{K}$ converges effectively to $u$. Therefore, $u$ is a computable continuous function.

Now, for the function $g(t) = \sum^{\infty}_{k=1} d_{k}\, p_{n_k}'(t)$, we have
\begin{equation*}
	\left\| g - P'_{K} \right\|_{\infty}
	\leq \textstyle\sum^{\infty}_{k=K+1} d_{k} \left\|p'_{n_k}\right\|_{\infty}
	= \textstyle\sum^{\infty}_{k=K+1} d_{k}
	= \textstyle\sum^{\infty}_{k=K+1} \left( r_{k} - r_{K-1} \right)
	= x - r_{K}
\end{equation*}
using $\left\|p'_{n}\right\|_{\infty} \leq 1$ for all $n\in\NN$ and  $\lim_{n\to\infty}r_{n} = x$.
So for $K\to\infty$, one gets $\lim_{K\to\infty} \left\| g - P'_{K} \right\|_{\infty} = 0$, showing that $g \in \C(\TT)$, and 
and Lemma~\ref{lem:AppxPoission} implies that $g = u'$, showing that $u \in \Uc$.
Finally, as in the proof of Theorem~\ref{thm:TuringMachine_f}, we have 
$\left| u'(t) \right| = \left| \textstyle\sum^{\infty}_{k=1} d_{k} p'_{n_k}(t) \right| \leq \textstyle\sum^{\infty}_{k=1} d_{k} = \left| u'(0) \right| = x$
for every $t\in\TT$, proving that $\left\|u'\right\|_{\infty} = x$.
\end{IEEEproof}

So with Theorem~\ref{thm:InfNormSigma1} and Theorem~\ref{thm:InfNorm_FktUc}, we immediately have the following statement.

\begin{corollary}
The mapping $\Op{T} : \Uc \to \RN_{+} = \left\{ x\in\RN : x>0 \right\}$ given by $\Op{T}(u) = \left\|u'\right\|_{\infty}$ satisfies $\Op{T}(\Uc) = \Sigma_{1} \cap \RN_{+}$.
\end{corollary}

According to Theorem~\ref{thm:InfNorm_FktUc}, to every $x \in \Sigma_{1}$ there exists a $u \in \Uc$ with $\left\|u'\right\|_{\infty} = x$.
It is an interesting question of practical importance whether it is possible to algorithmically construct this $u$ based on the given $x\in\Sigma_{1}$, i.e. does there exist a Turing machine with input $x$ and whose output is a function $u \in \Uc$ satisfying $\left\|u'\right\|_{\infty} = x$?
Analyzing the previous proof of Theorem~\ref{thm:InfNorm_FktUc}, we notice that the construction of $u\in\Uc$ was not constructive, i.e. the proof shows only the existence of such a $u$ but it gives no algorithm which allows us to construct $u$ from the given $x$.
In particular, there is no algorithmic way to obtain the sequence $\left\{ r_{n} \right\}_{n\in\NN}$, which was used in the proof to construct $u \in \Uc$.
However, if we restrict $x$ to lie in the interval $[0,1]$, then there exists such a Turing machine $\TM_{*}$ whose input is an algorithmic description of the input $x$ (namely the Zheng-Weihrauch representation of $x$) and whose output
is an approximation $\widetilde{u}$ of $u$ so that the approximation error can effectively be controlled.
This is shown by the following theorem.

\begin{theorem}
\label{thm:TuringMachine_f}
There exists a Turing machine $\TM_{*}$ with the following property: 
For an arbitrary $x \in \Sigma_{1} \cap [0,1]$ with Zheng-Weihrauch representation $\varphi_{A} : \NN \to A\subset \NN$,
the input of $\TM_{*}$ is $\varphi_{A}$ and the output is the function $u \in \Uc$ satisfying $\left\|u'\right\|_{\infty} = x$.
\end{theorem}

\begin{remark}
Remember, the output of $\TM_{*}$ is, in fact, an approximation $\widetilde{u} \in \Uc$, satisfying $\left\|u - \widetilde{u}\right\|_{\infty} < \epsilon$, and where the desired error bound $\epsilon>0$ is a further input to $\TM_{*}$.
\end{remark}

\begin{remark}
Note also, that $\TM_{*}$ is universal for this task in the sense that it does not depend on the given $x$.
\end{remark}

\begin{IEEEproof}
Let $x \in \Sigma_{1} \cap [0,1]$ be arbitrary with a corresponding recursively enumerable set $A\subset \NN$, and let $\varphi_{A} : \NN \to A$ be a Zheng-Weihrauch representation of~$x$.
Therewith, define the function 
\begin{equation*}
	u(t) = \textstyle\sum^{\infty}_{n=1} 2^{-\varphi_{A}(n)} p_{n}(t)
\end{equation*}
with $p_{n}$ given in \eqref{equ:trigPoly_pn}.
It follows from Theorem~\ref{thm:MainNegativResult} that $u \in \Uc$. 
Moreover, it was shown in \cite[Theorem~5]{BP_IEEECAS20} that
\begin{equation}
\label{equ:OutputTM}
	u_{m}(t) = \textstyle\sum^{m}_{n=1} 2^{-\varphi_{A}(n)} p_{n}(t)\;,
	\quad m\in\NN
\end{equation}
is a computable sequence $\left\{ u_{m} \right\}_{m\in\NN}$ of polynomials which satisfies
\begin{equation*}
	\lim_{m\to\infty} \left\| u - u_{m} \right\|_{\infty} = 0
	\quad\text{and}\quad
	\lim_{m\to\infty} \left\| u' - u'_{m} \right\|_{\infty} = 0\;.
\end{equation*}
Additionally, the definition of $p_{n}$ immediately implies $\left|p'_{n}(t)\right| \leq p'_{n}(0) = 1$ for all $t\in \TT$, so that 
\begin{equation*}
	\left| u'(t) \right|
	\leq \textstyle\sum^{\infty}_{n=1} 2^{-\varphi_{A}(n)} \left| p'_{n}(t) \right|
	\leq \textstyle\sum^{\infty}_{n=1} 2^{-\varphi_{A}(n)} p'_{n}(0)
	= u'(0)
	= \textstyle\sum^{\infty}_{n=1} 2^{-\varphi_{A}(n)}
	= x\;,
	\qquad \text{for all}\ t\in\TT\;.
\end{equation*}
This proves the statement of the theorem, because according to \eqref{equ:uA_um}, the approximation $u_{m}$ given in 
\eqref{equ:OutputTM} converges effectively to $u$.
\end{IEEEproof}

So \eqref{equ:OutputTM} gives the algorithm to effectively construct the desired $u\in\Uc$ from the given Zheng-Weihrauch representation $\varphi_{A}$ of~$x\in\Sigma_{1} \cap [0,1]$.

\section{Non-Dectability of Non-Computability of $u'$}
\label{sec:DecProblLTI}

Assume we want to use a digital computer to calculate $u'(0)$ for arbitrary functions $u\in\Uc$.
Since for some functions $u\in\Uc$, $u'(0)$ is not computable (cf. Theorem~\ref{thm:MainNegativResult}),
we have to verify beforehand whether the given $u\in\Uc$ has the property that $u'(0)$ is actually computable.
Otherwise the algorithm might calculate something, but it will not be possible to guarantee a predefined error bound so that we can't be sure whether the obtained result is sufficiently close to the true value $u'(0)$.
Therefore, it is desirable to have an exit-flag functionality as sketched in Fig.~\ref{fig:ProProc3}
and which is able to detect whether a given $u\in\Uc$ has the property that $u'(0)$ is computable or not.
If such a strong exit-flag functionality is not possible, 
we may ask for the weaker exit-flag functionality as shown in Fig.~\ref{fig:ProProc1} or Fig.~\ref{fig:ProProc2}.
They are only able to detect functions $u\in\Uc$ for which $u'(0)$ is computable or not computable, respectively (cf. also discussion in Section~\ref{sec:Indro}).

Concentrating first on the strong exit-flag functionality of Fig.~\ref{fig:ProProc3},
we ask whether it is possible to algorithmically decide, for any given $u\in\Uc$, whether $u'(0)\in\RNc$ or $u'(0)\notin\RNc$.
More precisely, we consider the following detection problem.

\begin{problem}[Decision problem]
\label{prob:GeneralDProblem}
Does there exist a Turing machine $\TMD(u)$ with input set $\Uc$ which halts for every $u\in\Uc$,
and with two output states
\begin{equation*}
	\TMD(u) = \left\{\begin{array}{lll}
	0 & \text{if} & u'(0)\notin\RNc\\
	1 & \text{if} & u'(0)\in\RNc
	\end{array}\right.\; ?
\end{equation*}
\end{problem}
A Turing machine $\TMD$ as in Problem~\ref{prob:GeneralDProblem} could be used to implement the strong exit-flag functionality sketched in Fig.~\ref{fig:ProProc3}.
If $\TMD(u)$ detects that $u'(0)\in\RNc$, then it will start the actual Turing machine $\TM_{u'}$ for computing $u'(0)$.
Otherwise, if $\TMD(u)$ detects that $u'(0)\notin\RNc$, then it stops the calculation with a corresponding information for the user.

From a mathematical point of view, the decision Problem~\ref{prob:GeneralDProblem} is equivalent to the question whether the set 
\begin{equation}
\label{equ:Ucomp}
	\Ucomp = \left\{ u \in \mathcal{U}_{\mathrm{c}}\ :\ u'(0) \in \RNc \right\}
\end{equation}
is decidable (cf. Def.~\ref{defi:DecidableSets}).
Equivalently, one may ask whether the complement set 
\begin{equation}
\label{equ:Unocomp}
	\Unocomp = \left\{ u \in \mathcal{U}_{\mathrm{c}}\ :\ u'(0) \notin \RNc \right\}
\end{equation}
is decidable.
Since $\Unocomp = \Ucomp^{\mathrm{c}} = \Uc\backslash\Unocomp$, it is clear that $\Ucomp$ is decidable (with respect to $\Uc$) if and only if $\Unocomp$ is decidable.

The following main result of this section shows that both sets, $\Ucomp$ and $\Unocomp$ are not only not decidable, but are also not semidecidable.

\begin{theorem}
\label{thm:DP_LTI}
Both set $\Ucomp$ and $\Unocomp$ as defined in \eqref{equ:Ucomp} and \eqref{equ:Unocomp}, respectively, are not semidecidable with respect to $\Uc$.
\end{theorem}

\begin{remark}
Note that if $\Ucomp$ is not semidecidable,
then this does not imply that $\Unocomp$ is not semidecidable, and vice versa.
This is different then for the decidability of the complementary sets $\Ucomp$ and $\Unocomp$.
So in the subsequent proof of Theorem~\ref{thm:DP_LTI} we have to prove both statements separately.
\end{remark}
Reformulating Theorem~\ref{thm:DP_LTI} in terms of the existence of particular Turing machines as explained in Section~\ref{sec:Decidability}, we get the following statement.
\begin{corollary}
\label{cor:DP_LTI}
There exist no Turing machines $\TM_{0}(u)$ and $\TM_{1}(u)$ with inputs $u\in\Uc$ both of which have only one output state, and so that
\begin{itemize}
\item $\TM_{0}(u)$ halts if and only if $u\in\Ucomp$
\item $\TM_{1}(u)$ halts if and only if $u\in\Unocomp$.
\end{itemize}
\end{corollary}
Corollary~\ref{cor:DP_LTI} immediately implies that the weak exit-flag functionalities sketched in Fig.~\ref{fig:ProProc1} and \ref{fig:ProProc2} cannot be implemented.
This shows that $\Ucomp$ has indeed much stronger non-computability properties than the semi-decidable set $\left\{ x\in\RNc : x\neq 0 \right\}$ shortly discussed in Example~\ref{exa:ExampleNonCompu} at the beginning of this paper.
Moreover, since $\Ucomp$ is not semidecidable, it is \emph{a fortiori} not decidable.
In particular, there exists no decider $\TMD(u)$ with input $u\in\Uc$ which is able to decide after finitely many iterations whether the input $u$ belongs to $\Ucomp$ or to $\Unocomp = \Ucomp^{\mathrm{c}} = \Uc\backslash\Ucomp$.
The same conclusion can be drawn from the fact that $\Unocomp$ is not decidable.

\begin{corollary}
There exists no Turing machine $\TMD(u)$ with input $u \in\Uc$ which always halts after finitely many iterations and with the two output states
\begin{equation*}
	\TMD(u) = 
	\left\{\begin{array}{lcl}
	0 & \text{if} & u \in \Unocomp\\
	1 & \text{if} & u\in\Ucomp\,.
	\end{array}\right.
\end{equation*}
\end{corollary}

Instead of Theorem~\ref{thm:DP_LTI}, we prove the equivalent statement of Corollary~\ref{cor:DP_LTI}.

\begin{IEEEproof}[Proof of Corollary~\ref{cor:DP_LTI}]
\emph{Part I:} We start with the statement for $\TM_{0}$.
To this end, let $u_{A}$ be the function \eqref{equ:function_uA} from Theorem~\ref{thm:MainNegativResult} and let $u_{0}$ be the zero function in $\Uc$, i.e.
$u_{0}(t) = 0$ for all $t\in\TT$.
For any $\lambda \in [0,1] \cap \RNc$, we define the function
\begin{equation*}
	u_{A,\lambda}(t) = \lambda u_{A}(t) + \left(1-\lambda\right) u_{0}(t)\,,
	\quad t\in \TT\;.
\end{equation*}
Since $u_{0}\in\Ucomp$, and according to Theorem~\ref{thm:MainNegativResult}, we have
\begin{equation}
\label{equ:ual_comp}
\begin{array}{lll}
	u_{A,\lambda} \in \Ucomp & \text{for} & \lambda = 0\\
	u_{A,\lambda} \in \Unocomp & \text{for all} & 0 < \lambda \leq 1\,.
\end{array}	
\end{equation}
In a similar way, we define for $\lambda \in [0,1] \cap \RNc$ the trigonometric polynomials
\begin{equation*}
	u_{m,\lambda}(t) = \lambda u_{m}(t) + \left(1-\lambda\right) u_{0}(t)\,,
	\quad t\in\TT\,,
\end{equation*}
with $u_{m}$ given in \eqref{equ:um}. 
Then \eqref{equ:uA_um} implies
\begin{equation}
\label{equ:UniformEffectiveConvergence}
	\left| u_{A,\lambda}(t) - u_{m,\lambda}(t) \right|
	\leq \frac{\lambda\, K_{0}}{G(m+1)}
	\leq \frac{K_{0}}{G(m+1)}
\end{equation}
for all $t\in\TT$ and for every $\lambda\in [0,1] \cap \RNc$.
So $u_{m,\lambda}$ converges uniformly and effectively to $u_{A,\lambda}$, independent of $\lambda\in [0,1]$.

After these preparations, we prove the statement of the corollary by contradiction and assume that there exists a Turing machine $\TM_{0}(u)$ with the properties given in Corollary~\ref{cor:DP_LTI}.
Moreover, $\TM_{>0}(\lambda)$ denotes a Turing machine with input $\lambda\in [0,1]\cap\RNc$, which halts if and only if $\lambda >0$, and which calculates forever if $\lambda = 0$.
Such a Turing machine does exist because the set $\left\{ \lambda\in [0,1]\cap\RNc\ :\ \lambda > 0\right\}$ is known to be semidecidable \cite{PourEl_Computability}.
Now we construct a Turing machine $\TMD(\lambda)$ with input $\lambda\in [0,1]\cap\RNc$ as follows:
\begin{enumerate}
\item[(i)]
For any input $\lambda\in [0,1]\cap\RNc$, $\TMD$ determines in a first step, $u_{A,\lambda}\in\Uc$. 
This is possible, because of \eqref{equ:UniformEffectiveConvergence}, i.e., because the sequence $\left\{u_{m,\lambda}\right\}_{m\in\NN}$ converges effectively and uniformly to $u_{A,\lambda}$, for all $\lambda\in [0,1]\cap\RNc$.
\item[(ii)]
Afterward, $\TMD$ starts the two Turing machines $\TM_{0}(u_{A,\lambda})$ and $\TM_{>0}(\lambda)$ in parallel. 
It halts if either $\TM_{0}$ or $\TM_{>0}$ halts, and it gives the output
\begin{equation*}
	\TMD(\lambda) = \left\{\begin{array}{ccl}
	\text{0} & \text{if} & \TM_{0}(u_{A,\lambda})\ \text{halts}\\
	\text{1} & \text{if} & \TM_{>0}(\lambda)\ \text{halts}.
	\end{array}\right.
\end{equation*}
\end{enumerate}
Because of \eqref{equ:ual_comp}, the Turing machine $\TM_{0}$ halts if and only if $\lambda = 0$, 
and by the definition of $\TM_{>0}$, the Turing machine $\TM_{>0}$ halts exactly if $\lambda > 0$.
So we constructed a Turing machine $\TMD : [0,1]\cap\RNc \to \left\{ 0, 1 \right\}$,
which always halts and which is able to decide whether the input $\lambda\in [0,1]\cap\RNc$ is equal to zero or larger than zero.
However, it is well known that such a Turing machine does not exist, because the set $\left\{ \lambda\in [0,1]\cap\RNc\ :\ \lambda > 0\right\}$ is not decidable, but is only semidecidable \cite{PourEl_Computability}. 
So we have obtained a contradiction. 
Therefore, the assumption that $\TM_{0}(u)$ exists was wrong, which proves the first statement of the corollary.

\emph{Part II}: 
Also, the proof of the second statement, concerning the existence of $\TM_{1}$, is based on the function $u_{A}$ defined by \eqref{equ:function_uA} in Theorem~\ref{thm:MainNegativResult}.
Then for any $\rho \in [0,1)$, we consider the Poisson integral $u_{A}(\rho,t) = \left( \Op{P}_{\rho} u_{A} \right)(t)$ of $u_{A}$ given by \eqref{equ:PoissonIntegral} for $t\in\TT$. 
By known properties of the Poisson integral \cite{Garnett}, we have that $u_{A}(\rho,\cdot) \in \C^{\infty}(0,2\pi)$ for every $\rho\in [0,1)$ and $u_{A}(1,\cdot) = u_{A}$.

From the definition in \eqref{equ:function_uA} and \eqref{equ:trigPoly_pn}, we note that $u_{A}$ can be written as a sine series
\begin{equation}
\label{equ:SinSeries}
	u_{A}(t) = \textstyle\sum^{\infty}_{k=2} b_{k}\, \sin(k t)\,,
\end{equation}
with a coefficient sequence $\left\{ b_{k} \right\}^{\infty}_{k=2}$ whose concrete values are not important here.
This representation of $u_{A}$ shows that the corresponding Poisson integral 
is given by
\begin{equation*}
	u_{A}(\rho,t) = \textstyle\sum^{\infty}_{k=2} b_{k}\, \rho^{k}\, \sin(k t)\,.
\end{equation*}
Similarly as for $u_{A}$, we write the partial sums \eqref{equ:um} of $u_{A}$ as the sine series 
\begin{equation*}
	u_{m}(t) = \textstyle\sum^{m+1}_{k=2} b_{k}(m)\, \sin(k t)\;,
	\quad t\in\TT\,,
\end{equation*}
where the coefficients $\left\{ b_{k}(m) \right\}^{m+1}_{k=2}$ depend now on the degree $m$.
The corresponding Poisson integral of $u_{m}$ is then given by
\begin{equation*}
	u_{m}(\rho,t)
	= \left( \Op{P}_{\rho} u_{m} \right)(t)
	= \textstyle\sum^{m+1}_{k=2} b_{k}(m)\, \rho^{k}\, \sin(k t)\,.
\end{equation*}
This shows that $\left\{ u_{m}(\cdot,t)\right\}^{\infty}_{m=2}$ is a computable sequence of polynomials for every fixed $t\in\TT$ 
and that $\left\{ u_{m}(\rho,\cdot)\right\}^{\infty}_{m=2}$ is a computable sequence of trigonometric polynomials for every fixed $\rho\in [0,1)$.
By the maximum modulus principle \eqref{equ:maxModulusPoission} for harmonic functions, we get from \eqref{equ:uA_um}
\begin{equation}
\label{equ:EffConv}
	\left| u_{A}(\rho,t) - u_{m}(\rho,t) \right|
	= \frac{1}{2\pi}\int^{2\pi}_{0} \left| u_{A}(\tau) - u_{m}(\tau) \right|\, \mathcal{P}_{\rho}(t-\tau)\, \d\tau
	\leq \max_{\tau\in [0,2\pi)} \left| u_{A}(\tau) - u_{m}(\tau) \right|
	\leq \frac{K_{0}}{G(m+1)}
\end{equation} 
for all $\rho\in [0,1)\cap\RNc$, using that $\int^{2\pi}_{0}\mathcal{P}_{\rho}(\tau)\, \d\tau = 2\pi$ for every $\rho\in[0,1)$.
Thus $u_{m}(\rho,t)$ effectively converges to $u_{A}(\rho,t)$ as $m\to\infty$ uniformly for all $t \in \TT$ and all $\rho\in [0,1)\cap\RNc$.

 So from the previous definition of the functions $u_{A}(\rho,\cdot)$, we see that 
\begin{align*}
\begin{array}{rl}
	i) & u_{A}(\rho,\cdot) \in \Cac(\TT)\\
	ii) & u'_{A}(\rho,\cdot) \in \W\\
	iii) & u'_{A}(\rho,0) \in \RNc
\end{array}	
\end{align*}
for all $\rho\in [0,1)\cap\RNc$.
Indeed, Property~$i)$ is obvious from $u_{A}(\rho,\cdot) \in \C^{\infty}(\TT)$.
Property~$ii)$ follows from
\begin{equation*}
	\left\| u'_{A}(\rho,\cdot) \right\|_{\W}
	= \textstyle\sum^{\infty}_{k=2} \left|k\right|\, \left|b_{k} \right|\, \rho^{k}
	\leq \left\| u'_{A} \right\|_{\W} < \infty\;,
\end{equation*}
and Property~$iii)$ is a consequence of the effective convergence noted in \eqref{equ:EffConv}.
Overall, we have shown that $u_{A}(\rho,\cdot) \in \Uc$ for all $\rho\in [0,1]\cap\RNc$ and 
\begin{equation}
\label{equ:uarho_comp}
\begin{array}{lll}
	u_{A}(\rho,\cdot) \in \Ucomp & \text{for} & 0\leq\rho < 1\\
	u_{A}(\rho,\cdot) \in \Unocomp & \text{for} &  \rho = 1.
\end{array}	
\end{equation}

Now we proceed very similarly as in the first part of the proof and assume, in contradiction to the statement of the corollary, that there exists a Turing machine $\TM_{1}$ with the properties claimed by the corollary.
Moreover, let $\TM_{<1}(\rho)$ be a Turing machine with input $\rho\in [0,1]\cap\RNc$ and which halts after finitely many iterations if and only if $\rho < 1$. Otherwise, if $\rho = 1$, the Turing machine will never halt.
Such a Turing machine does exist because the subset $\left\{ \rho\in [0,1]\cap\RNc : \rho < 1 \right\}$ of $[0,1]\cap\RNc$ is known to be semidecidable \cite{PourEl_Computability}. Then we construct a Turing machine $\TMD(\rho)$ with input $\rho\in [0,1]\cap\RNc$ as follows:
\begin{enumerate}
\item[(i)]
For any input $\rho\in [0,1]\cap\RNc$, the Turing machine $\TMD$ first determines the function $u_{A}(\rho,\cdot)\in\Uc$.
\item[(ii)] Afterward, $\TMD$ starts the Turing machines $\TM_{<1}(\rho)$ and $\TM_{1}(u_{A}(\rho,\cdot))$ in parallel, and it halts if either $\TM_{<1}$ or $\TM_{1}$ halts with the output
\begin{equation*}
	\TMD(\rho)
	= \left\{ \begin{array}{ccl}
	\text{0} & \text{if} & \TM_{<1}(\rho)\ \text{halts}\\
	\text{1} & \text{if} & \TM_{1}(u_{A}(\rho,\cdot))\ \text{halts.}
	\end{array}\right.	
\end{equation*}
\end{enumerate}
$\TM_{<1}(\rho)$ halts if and only if $\rho < 1$
and because of \eqref{equ:uarho_comp}, $\TM_{1}$ halts if and only if $\rho = 1$.

This way, we constructed a decider, i.e. the Turing machine
$\TMD : [0,1]\cap\RNc \to \left\{ 0 , 1 \right\}$
which always halts and which is able to decide whether a given $\rho\in [0,1]\cap\RNc$ is smaller than $1$ or equal to $1$.
However, it is known that such a Turing machine does not exist because the subset $\left\{ \rho \in [0,1]\cap\RNc : \rho<1\right\}$ of $[0,1]\cap\RNc$ is not decidable but only semidecidable \cite{PourEl_Computability}.
So we arrive at a contradiction.
Therefore our assumption that $\TM_{1}$ exists was wrong, proving the statement of the corollary.
\end{IEEEproof}

\section{Non-Detectability of Upper Bounds for $\left\|u'\right\|_{\infty}$}
\label{sec:DetectionUpperBound}

As discussed in the introduction, there are many applications where it is not necessary to know $u'(t)$ at a specific $t\in\RNc$, but it would be sufficient to know $\left\| u' \right\|_{\infty}$, or at least a good upper bound for $\left\| u' \right\|_{\infty}$.
Nevertheless, it was shown in Section~\ref{sec:ComputeFirstDerivatice} that for $u \in \Uc$, the maximum norm $\left\|u'\right\|_{\infty}$ is generally not algorithmically computable.
Therefore, we investigate now whether it is possible to detect at least an upper bound for $\left\|u'\right\|_{\infty}$.
It will be shown that this is not possible and we discuss consequences for recent quantum based algorithms. 

Assume $u \in \Uc$ and a positive computable number $\lambda$ is given.
Is it possible to decide algorithmically (i.e. by a Turing machine) whether $\lambda$ is an upper bound for $\left\|u'\right\|_{\infty}$?
More formally, we investigate the following problem.

\begin{defi}
\label{def:AlgorithmicSolvable}
Let $u\in\Uc$ be arbitrary.
We say that the detection of an upper bound for $\left\|u'\right\|_{\infty}$ is algorithmically solvable if there exists a Turing machine $\TM_{*}$ whose input is a computable number $\lambda > 0$ and which stops if and only if $\lambda > \left\|u'\right\|_{\infty}$. Otherwise, it never halts.
\end{defi}
Note that even through $\TM_{*}$ generally depends on $u$, it is not required that $\TM_{*}$ depends recursively (i.e. algorithmically) on the given function $u$. So $\TM_{*}$ might be a Turing machine which was specifically designed (by a human) for the particular but arbitrary $u\in\Uc$. 
Moreover, it is only required that $\TM_{*}$ stops if the input is an upper bound for $\left\|u'\right\|_{\infty}$.
For other inputs, it calculates forever.
In other words, we ask whether for an arbitrary given $u\in\Uc$, the set
\begin{equation*}
	\mathcal{R}_{u} = \left\{ x \in \RNc : \left\|u'\right\|_{\infty} < x \right\}
\end{equation*}
is semidecidable.
The following theorem shows that it is not, and it precisely classifies those $u\in\Uc$ for which $\mathcal{R}_{u}$ is not semidecidable.

\begin{theorem}
\label{thm:DetectingUpperBound}
Let $u\in\Uc$ with $\left\|u'\right\|_{\infty} \in \Sigma_{1}\backslash \Delta_{1}$, arbitrary.
Then the problem of detecting an upper bound for $\left\|u'\right\|_{\infty}$, as described by Def.~\ref{def:AlgorithmicSolvable}, is not algorithmically solvable.
\end{theorem}

\begin{IEEEproof}
Assume the opposite, namely that there exists an $u\in\Uc$ with $\left\|u'\right\|_{\infty} \in \Sigma_{1}\backslash \Delta_{1}$, so that for this function the problem of detecting an upper bound for $\left\|u'\right\|_{\infty}$ is algorithmically solvable.
We denote by $\TM_{*}$ the corresponding Turing machine, and we define for $i\in\NN$ the dyadic rational numbers $\lambda_{i,k} = k\cdot 2^{-i}$, $k=1,\dots,2^{i}-1$.

First we assume that $\left\|u'\right\|_{\infty} < \lambda_{0} = 1$, and describe a procedure to find an upper bound for $\left\|u'\right\|_{\infty}$.
We start the Turing machine $\TM_{*}$ with input $\lambda_{1,1} = 1/2$ and perform the first calculation step of $\TM_{*}(\lambda_{1,1})$.
If, after this first step, $\TM_{*}(\lambda_{1})$ already stops, we have found an upper bound $\lambda_{1} = 1/2$.
Otherwise, if $\TM_{*}(\lambda_{1,1})$ does not stop after the first step, we calculate the second step of $\TM_{*}(\lambda_{1,1}) = \TM_{*}(\lambda_{2,2})$ and in parallel, we calculate the first steps of $\TM_{*}(\lambda_{2,1})$ and $\TM_{*}(\lambda_{2,3})$.
If one of these three Turing machines stops after this step, we have found a corresponding upper bound $\lambda_{1}$ for $\left\|u'\right\|_{\infty}$, namely the smallest input $\lambda_{2,k}$ among all Turing machines which stopped in this second step.
If no Turing machine has stopped, we calculate the next step of the running Turing machines, and in parallel, we start Turing machines with inputs $\lambda_{3,1}$, $\lambda_{3,3}$, $\lambda_{3,5}$, $\lambda_{3,7}$.
Again, if at least one of the Turing machines stops after this step, we have found an upper bound $\lambda_{1}$ for $\left\|u'\right\|_{\infty}$, otherwise we continue with step $4$.

Since $\left\|u'\right\|_{\infty} < 1$ and $\TM_{*}(\lambda)$ stops for all $\lambda > \left\|u'\right\|_{\infty}$, it is clear that the above procedure stops after finitely many steps with a dyadic rational number $\lambda_{1} < 1$, as the upper bound for $\left\|u'\right\|_{\infty}$.
So in particular, we found that $\left\|u'\right\|_{\infty} < \lambda_{1}$. 
Next, we again apply the above procedure, but start now with the assumption $\left\|u'\right\|_{\infty} < \lambda_{1}$.
This yields another dyadic rational number $\lambda_{2} < \lambda_{1}$ as an upper bound for $\left\|u'\right\|_{\infty}$,
i.e we find $\left\|u'\right\|_{\infty} < \lambda_{2}$.
Now we apply this procedure iteratively to obtain a computable sequence $\left\{ \lambda_{n} \right\}_{n\in\NN} \subset [0,1)$ of dyadic rational numbers, so that $\lambda_{n+1} < \lambda_{n}$ and $\left\|u'\right\|_{\infty} < \lambda_{n}$ for all $n\in\NN$.
Consequently, there exists
\begin{equation}
\label{equ:LimitLambda}
	\lambda_{\infty} = \textstyle\lim_{n\to\infty} \lambda_{n}
\end{equation}	
which satisfies $\left\|u'\right\|_{\infty} \leq \lambda_{\infty}$.
Next we show that actually, $\left\|u'\right\|_{\infty} = \lambda_{\infty}$.
Indeed, assume $\left\|u'\right\|_{\infty} < \lambda_{\infty}$.
Then there exists a dyadic rational number $\lambda_{*}$ so that $\left\|u'\right\|_{\infty} < \lambda_{*} < \lambda_{\infty}$.
But this would imply that $\lambda_{*} = \lambda_{i,k}$ was an input in the above described algorithm in step $i$ and in all following steps.
Consequently $\TM(\lambda_{*})$ must have stopped in this previous algorithm, and so there exists an $n_{0}$ so that $\lambda_{n} < \lambda_{*}$ for all $n > n_{0}$.
But this contradicts \eqref{equ:LimitLambda} and we necessarily have $\left\|u'\right\|_{\infty} = \lambda_{\infty} = \lim_{n\to\infty} \lambda_{n}$.
So, $\left\|u'\right\|_{\infty}$ is the limit of a monotone decreasing computable sequence $\left\{ \lambda_{n} \right\}_{n\in\NN}$.
However, since $\left\|u'\right\|_{\infty} \in \Sigma_{1}$, it is also the limit of a monotone increasing computable sequence,
i.e. there exists a computable sequence $\left\{ r_{n} \right\}_{n\in\NN}$ of rational numbers with $r_{n} \leq r_{n+1}$ for all $n\in\NN$ and with $\left\| u' \right\|_{\infty} = \lim_{n\to\infty} r_{n}$.
Consequently, it follows that $\left\|u'\right\|_{\infty} \in \RNc = \Delta_{1}$.
However, this contradicts the assumption that $\left\|u'\right\|_{\infty} \in \Sigma_{1}\backslash \Delta_{1}$.
So the Turing machine $\TM_{*}$ does not exist.
\end{IEEEproof}

Theorem~\ref{thm:DetectingUpperBound} has practical implications in the theory of quantum computing. 
There the so-called hidden subgroup problem plays an important role, and recently several quantum algorithms were proposed for solving this problem \cite{Eisentraeger_STOC14,BiasseSong_QuantumAlgorithm2016,BoerDucasFehr_QuantumComplex20}.
These algorithms are generalizations of \emph{Shor's} famous algorithm \cite{Shor_SIAM99} for breaking \emph{Rivest--Shamir--Adleman} (RSA) and \emph{Diffie--Hellman-Merkle} (DHM) cryptosystems,
and generalizations of the \emph{Hallgren's} algorithm \cite{Hallgren_ACM07} for solving \emph{Pell's equation}, and therefore for breaking the \emph{Buchmann--Williams key-exchange} cryptosystem.
In particular, these algorithms are much more powerful than Shor's or Hallgren's algorithm and they even allow us to break several lattice-based cryptosystems.
Therefore, they provide very interesting approaches to attack so-called post-quantum cryptography methods.
However, in contrast to Shor's algorithm, the approaches in \cite{Eisentraeger_STOC14,BiasseSong_QuantumAlgorithm2016,BoerDucasFehr_QuantumComplex20} explicitly need $\left\|u'\right\|_{\infty}$ as an input.
Without this input, no bounds on the complexity in form of the necessary quantum operations can be derived.
Consequently, it would not be possible to claim an exponential speedup of these quantum algorithms compared to the best known classical algorithm.
So Theorem~\ref{thm:DetectingUpperBound} points to a limitation of these algorithms since it shows that even for a fixed $u\in\Uc$ the problem of detecting an upper bound for $\left\|u'\right\|_{\infty}$ is not algorithmically solvable.
Consequently, such an upper bound can generally not be found algorithmically, but has to be determined explicitly by mathematical methods.
Finally, it may be interesting to note that Shor's algorithm is basically an algorithm for finding the period of a function.
Using the concrete structure of the prime factorization problem and properties of the quantum Fourier transform, Shor was able to obtain all necessary information about the variation of the function for which the period has to be determined.
For more general problems, such a structure is not yet known \cite{Shor_ACM03,Shor_QIP04}.

\section{Non-Detectability for the Wave Equation}
\label{sec:WaveEquation}

This section investigates the detectability of the non-computability of solutions of the wave equation by observing the initial values.
We start by giving a short review of the Cauchy problem for the wave equation in $\RN^{3}$ and recall the results showing that there exist initial values such that the solution is not computable at certain points.
Afterward, it is proved that this non-computability is not detectable by a Turing machine.

\subsection{The Cauchy problem of the wave equation}

We write $x = (x_{1},x_{2},x_{3}) \in \RN^{3}$ for a vector in $\RN^{3}$ with its three Cartesian coordinates $x_{1},x_{2},x_{3}\in\RN$ and with its usual norm $\|x\| = \left( x^{2}_{1} + x^{2}_{2} + x^{2}_{3} \right)^{1/2}$.
For any twice partially differentiable function $u : \RN_{+} \times \RN^{3} \to\RN$, 
the \emph{Laplace operator} $\Delta$ is defined for $(t,x)\in \RN_{+} \times \RN^{3}$ by
\begin{equation*}
	\left( \Delta u \right)(t,x)
	= \frac{\partial^{2} u}{\partial x^{2}_{1}}(t,x) + \frac{\partial^{2} u}{\partial x^{2}_{2}}(t,x) + \frac{\partial^{2} u}{\partial x^{2}_{3}}(t,x)\,,	
\end{equation*}
and $u_{t} = \frac{\partial u}{\partial t}$ and $u_{tt} = \frac{\partial^{2} u}{\partial t^{2}}$ stand for the first and second partial derivative with respect to the time coordinate, respectively.
Then we consider the following initial value problem of the three-dimensional wave equation
\begin{equation}
\label{equ:WaveEquation}
\begin{array}{lll}
	u_{tt}(t,x)= c^{2} \left( \Delta u \right)(t,x)\,, &   & t>0\,, x\in\RN^{3},\\[0.618ex]
	u(0,x) = f(x)\,,\quad u_{t}(0,x) = 0\,, &  & x\in\RN^{3}.
\end{array}	
\end{equation}
Therein $c \in \RNc$, $c>0$ characterizes the velocity of the wave described by this differential equation. 
Without loss of generality, we assume $c = 1$.
The function $f$, i.e. the initial condition of the differential equation, describes the waveform at the initial time $t=0$.
Assuming $f \in \C^{1}(\RN^{3})$, the unique solution of \eqref{equ:WaveEquation} is known \cite{PouEL_MathLogQuart97,WeihrauchZhong_ProcLMathSoc02} to be
\begin{equation}
\label{equ:solution_u1}
	u(t,x)
	= \left( \Op{W} f\right)(t,x)
	= \frac{\partial}{\partial t} \Bigg( \frac{1}{4\pi t} \!\!\!\int\limits_{\partial B(x,t)}\!\!\! f(y)\, \d F(y) \Bigg)
\end{equation}
for $t>0$ and every $x\in\RN^{3}$.
Therein, one has to integrate over the sphere
$\partial B(x,t) = \left\{ \xi\in\RN^{3} : \left\| \xi - x \right\| = t \right\}$ in $\RN^{3}$ with center $x\in\RN^{3}$ and radius $t > 0$.
The surface measure on the three-dimensional sphere is denoted by $\d F$ and we write $\Op{W} : f \mapsto u$ for the operator mapping the initial conditions $f\in\C^{1}(\RN^{3})$ onto the corresponding solution $u$ of \eqref{equ:WaveEquation}.

\subsection{Computability of solutions of the wave equation}

As for the ordinary differential operator in Section~\ref{sec:ComputeFirstDerivatice}, we ask whether it is true that every computable input $f$ yields a solution $u$ of \eqref{equ:WaveEquation} which is again computable.
It is known that the answer is negative. 
In particular, \cite{PouEL_MathLogQuart97} constructed a computable function $f$ such that corresponding solution $u$ of \eqref{equ:WaveEquation} is not computable.
In this paper we use a significantly sharper version of this statement taken from \cite{BP_IEEECAS20}.
To this end, we define first the set
\begin{equation*}	
	\Vc 
	= \left\{\begin{array}{l}
	f  \in \CC(\RN^{3}) \cap \C^{1}(\RN^{3})\ \text{with the properties}\\[0.618ex] 
	\quad \bullet\ \frac{\partial f}{\partial x_{n}} \in \Cac(\RN)\ \text{for}\ n=1,2,3\\
	\quad \bullet\ f(x) = 0\ \text{for}\ \left\|x\right\| < \pi\ \text{and}\ \left\|x\right\| > 3\pi
	\end{array}\right\}
\end{equation*}
of all computable continuous functions, supported on a spherical shell in $\RN^{3}$, which are continuously differentiable and for which the partial derivatives (considered as functions of one variable $x_{n}$ while the other two variables are kept fixed) are absolutely continuous.
It was shown in \cite{BP_IEEECAS20} that the set $\Vc$ contains functions $f$ such that the corresponding solution $u = W(f)$ of the Cauchy problem \eqref{equ:WaveEquation} is not a computable function.
More precisely, the following statement was proven.

\begin{theorem}
\label{thm:WaveEquation}
There exists an $f\in\Vc$ such that the corresponding solution \eqref{equ:solution_u1} of the initial value problem \eqref{equ:WaveEquation} has the property that $u(2\pi,0) = \left(\Op{W} f\right)(2\pi,0)\notin \RNc$.
\end{theorem}
Theorem~\ref{thm:WaveEquation} shows that there are computable initial conditions $f\in \CC(\RN^{3}) \cap \C^{1}(\RN^{3})$ with absolutely continuous first partial derivatives
so that the solution $u = \Op{W}f$ of the wave equation \eqref{equ:WaveEquation} is not computable at position $x = (0,0,0)\in\RNc^{3}$ and at time $t = 2\pi\in\RNc$.

Similarly as in Section~\ref{sec:ComputeFirstDerivatice}, we decompose $\Vc$ into two complementary subsets 
\begin{equation}
\label{equ:Vsets}
\begin{split}
	\Vcomp &= \left\{ f\in\Vc\ :\ \left(\Op{W}f\right)(2\pi,0) \in \RNc \right\}\\
	\Vnocomp &= \left\{ f\in\Vc\ :\ \left(\Op{W}f\right)(2\pi,0) \notin \RNc \right\}
\end{split}	
\end{equation}
of all those initial values $f\in\Vc$, which yield a solution that is computable at $x = (0,0,0)\in\RNc^{3}$ and $t = 2\pi\in\RNc$, and all those that yield a solution which is not computable at this point, respectively.
It is clear from \eqref{equ:Vsets} that $\Vnocomp = \Vcomp^{\mathrm{c}} = \Vc\backslash\Vcomp$ and we are going to show in the next subsection that both sets are not semidecidable.
To this end, it will be important that the function $f\in\Vc$ in Theorem~\ref{thm:WaveEquation} has a particular structure.
Consider the set
\begin{equation}
\label{equ:SetMc}
	\Mc
	= \left\{\begin{array}{l}
	q\in\CC(\RN)\cap\C^{1}(\RN)\ \text{with the properties}\\[0.618ex]
	\quad \bullet\ q'\in\Cac(\RN)\\
	\quad \bullet\ q(t) = 0\ \text{for}\ t<\pi\ \text{and}\ t>3\pi
	\end{array}\right\}\,.
\end{equation}
The proof of Theorem~\ref{thm:WaveEquation} in \cite{BP_IEEECAS20} constructed a $q \in \Mc$ such that the corresponding function $f_{q} \in \Vc(\RN^{3})$, defined by
\begin{equation}
\label{equ:f_von_g}
	f_{q}(x) = q\left(\left\|x\right\|\right)
	= q\left(\sqrt{x^{2}_{1} + x^{2}_{2} + x^{2}_{3}}\right)\,,
	\quad x\in\RN^{3}\;,
\end{equation}
has the properties claimed in Theorem~\ref{thm:WaveEquation}. 
For any $f_{q}\in\Vc$, constructed as in \eqref{equ:f_von_g}, 
the closed-form solution of the wave equation \eqref{equ:solution_u1} is given for $t>0$ by
\begin{align}
\label{equ:ut0}
	u(t,0) &= \left(\Op{W} f_{q}\right)(t,0)
	= \frac{\partial}{\partial t} \bigg( \frac{1}{4\pi t} \int_{\partial B(0,t)}\!\!\! f(y)\, \d F(y) \bigg)
	= \frac{\partial}{\partial t} \bigg( \frac{1}{4\pi t} \int_{\left\| y \right\| = t}\!\!\! q(t)\, \d F(y) \bigg)
	= \frac{\partial}{\partial t} \bigg( \frac{q(t)}{4\pi t}\cdot 4\pi t^{2} \bigg)\nonumber\\
	&= q(t) + t q'(t)\;.
\end{align}
In particular, we have $\left(\Op{W} f_{q}\right)(2\pi,0) = q(2\pi) + 2\pi\, q'(2\pi)$.
Since $q\in\Mc$ and because $2\pi\in\RNc$, we thus have
\begin{equation*}
	\left(\Op{W}f_{q}\right)(2\pi,0) \in \RNc
	\quad\text{if and only if}\quad
	q'(2\pi)\in\RNc;.
\end{equation*}
This motivates us to define the following complementary subsets of $\Mc$,
\begin{align*}
	\Mcomp &= \left\{ q\in\Mc\ :\ q'(2\pi)\in\RNc \right\}\\
	\Mnocomp &= \left\{ q\in\Mc\ :\ q'(2\pi)\notin\RNc \right\}\;.
\end{align*}
Therefore, the previous discussion can be summarize by the following statement:

\begin{lemma}
\label{lem:McompVcomp}
Let $f_{q}\in\Vc$ be a function of the form \eqref{equ:f_von_g} with $q \in \Mc$.
Then $f_{q} \in \Vcomp$ if and only if $q \in \Mcomp$,
and conversely $f_{q} \in \Vnocomp$, if and only if $q \in \Mnocomp$.
\end{lemma}

\subsection{Detection problems for the wave equation}

Similarly as in Section~\ref{sec:ComputeFirstDerivatice}, we are going to show that there exists no decider, i.e. no algorithm which can be implemented on a Turing machine and which is always able to decide after a finite number of iterations whether a given function $f \in \Vc$ belongs to $\Vcomp$ or $\Vnocomp$.
So there is no decider which can detect whether $f\in\Vc$
yields a solution \eqref{equ:solution_u1} of the Cauchy problem \eqref{equ:WaveEquation} which is computable at $x=0$ and $t=2\pi$ or whether it yields a solution which is not computable at $x=0$ and $t=2\pi$.
If such a decider would exist, this would imply that $\Vcomp$ and $\Vnocomp$ are decidable sets.
However, we can prove that both sets are not even semidecidable.

\begin{theorem}
\label{thm:Vcomp}
Both sets $\Vcomp$ and $\Vnocomp$, defined in \eqref{equ:Vsets}, are not semidecidable.
\end{theorem}
Theorem~\ref{thm:Vcomp} may be reformulated in terms of the existence of a Turing machine with only one output state as explained in Section~\ref{sec:Decidability}.
This yields the following equivalent reformulation of Theorem~\ref{thm:Vcomp}.

\begin{corollary}
\label{cor:WaveEqu}
There exist no Turing machines $\TM_{0}(f)$ and $\TM_{1}(f)$ with inputs $f\in\Vc$, in which each has only one output state, and so that
\begin{itemize}
\item $\TM_{0}(f)$ halts if and only if $f\in\Vcomp$
\item $\TM_{1}(f)$ halts if and only if $f\in\Vnocomp$.
\end{itemize}
\end{corollary}
Moreover, since both subsets $\Vcomp$ and $\Vnocomp$ are not semidecidable they are \emph{a fortiori} not decidable.
Consequently, we immediately have the following observation.
\begin{corollary}
There exists no Turing machine $\TMD$ with input $f \in\Vc$, which always halts after finitely many iterations, and with the two output states
\begin{equation*}
	\TMD(f) = 
	\left\{\begin{array}{lcl}
	0 & \text{if} & f \in \Vnocomp\\
	1 & \text{if} & f\in\Vcomp\,.
	\end{array}\right.
\end{equation*}
\end{corollary}

Similarly as in Section~\ref{sec:DecProblLTI}, we prove Corollary~\ref{cor:WaveEqu} instead of the equivalent statement of Theorem~\ref{thm:Vcomp}.
The idea of the proof is very similar to the proof of Corollary~\ref{cor:DP_LTI}. 
We mainly have to adapt it to the particular situation of the wave equation.

\begin{IEEEproof}[Proof of Corollary~\ref{cor:WaveEqu}]
\emph{Part I:}
We prove first the non-existence of the Turing machine $\TM_{0}$.
To this end, we notice that there exists a $q \in \Mnocomp \subset \Mc$ so that the function $f_{q}$ defined by \eqref{equ:f_von_g} belongs to $\Vnocomp$ (cf. Lemma~\ref{lem:McompVcomp}).
Based on this $f_{q}$, we define for any $\lambda \in [0,1]\cap\RNc$ the function
\begin{equation*}
	f_{\lambda} = \lambda\, f_{q} + (1-\lambda)\, f_{0}
\end{equation*}
with the zero function $f_{0}$ of $\Vc$, i.e. $f_{0}(x) = 0$ for all $x\in\RN^{3}$.
By this definition, $f_{\lambda}\in\Vc$ for every $\lambda \in [0,1]\cap\RNc$ and since $f_{0}\in\Vcomp$, we have
\begin{equation}
\label{equ:flambdaFaelle}
\begin{array}{lll}
	f_{\lambda} \in \Vcomp & \text{if} & \lambda = 0\\
	f_{\lambda} \in \Vnocomp & \text{for all} & 0 < \lambda\leq 1\;.
\end{array}	
\end{equation}

Now we assume in contradiction to the statement of the corollary that there exists a Turing machine $\TM_{0}(f)$ with input $f \in \Vc$ and which halts if and only if the input $f$ belongs to $\Vcomp$.
Otherwise, if $f \in \Vnocomp$, it runs forever.
Moreover, let $\TM_{>0}(\lambda)$ be the Turing machine with input $\lambda\in [0,1]\cap\RNc$ which is defined as in Part~I of the proof of Corollary~\ref{cor:DP_LTI}.
Then we construct a Turing machine $\TMD(\lambda)$ with input $\lambda\in [0,1]\cap\RNc$ as follows:
\begin{enumerate}
\item[(i)]
For the input $\lambda$, $\TMD$ determines first the function $f_{\lambda}$.
This is possible because $f_{\lambda}\in\Vc$ is a computable continuous function.
\item[(ii)]
In a second step, $\TMD$ starts the two Turing machines $\TM_{0}(f_{\lambda})$ and $\TM_{>0}(\lambda)$ in parallel.
The Turing machine stops if either $\TM_{0}$ or $\TM_{>0}$ stops and gives the output
\begin{equation*}
	\TMD(\lambda) = 
	\left\{\begin{array}{lll}
	0 & \text{if} & \TM_{0}(f_{\lambda})\ \text{halts}\\[0.618ex]
	1 & \text{if} & \TM_{>0}(\lambda)\ \text{halts}.
	\end{array}\right.
\end{equation*} 
\end{enumerate}
Because of \eqref{equ:flambdaFaelle}, $\TM_{0}$ halts if and only if $\lambda = 0$, and by its definition, $\TM_{>0}$ halts if and only if $\lambda > 0$.
In other words, we constructed a decider, i.e. a Turing machine $\TMD$ that always halts for any input $\lambda\in [0,1]\cap\RNc$ and which is able to decide whether $\lambda = 0$ or $\lambda>0$.
Nevertheless, such a Turing machine does not exist because the subset $\left\{ \lambda \in [0,1]\cap\RNc : \lambda > 0\right\}$ of $[0,1]\cap\RNc$ is known to be not decidable \cite{PourEl_Computability}.
So we arrived at a contradiction, proving that the Turing machine $\TM_{0}$ does not exist. Thus the first statement of the corollary is verified.

\emph{Part II:} To prove the second statement, let $q\in\Mnocomp$ be arbitrary. 
Since $q$ is a computable continuous function on $[\pi,3\pi]$,
there exists (cf. Def.~\ref{defi:ComputableFkt}) a sequence of trigonometric polynomials $\left\{ p_{m} \right\}_{m\in\NN}$ and a recursive function $e : \NN\to\NN$, so that for every $N\in\NN$ and for all $t\in [\pi, 3\pi]$,
\begin{equation}
\label{equ:ApproxByPm}
	\left| q(t) - p_{m}(t) \right| \leq 2^{-N}
	\quad\text{for all}\ m\geq e(N)\,.
\end{equation}
Next we consider for $\rho\in (0,1]\cap\RNc$ the Poisson integral on the upper half plain of $q$ given by
\begin{equation*}
	q(\rho,t) 
	= \frac{1}{\pi}\int^{3\pi}_{\pi} q(\tau)\, \frac{\rho}{\rho^{2} + (t-\tau)^{2}}\, \d\tau\;,
	\quad t\in\RN\,,
\end{equation*}
and similarly, $p_{m}(\rho,t)$ stands for the Poisson integral of the trigonometric polynomial $p_{m}$.
By properties of the Poisson integral \cite{Garnett}, it follows that $q(\rho,\cdot) \in \C^{\infty}(\RN)$ for every $\rho\in (0,1]$, and since $q\in\CC(\RN)$, it follows that $q^{(k)}(\rho,\cdot) \in \CC(\RN)$ for every $k\in\NN$.
Moreover, for $\rho = 0$, we have $q(0,t) = q(t)$ for all $t\in\RN$.

\begin{figure}[t]
\begin{center}
\begin{picture}(150,60)(-10,-5)
    \put(-10, 0){\vector(1,0){150}}
		\put(   0,-5){\vector(0,1){55}}
		\multiput( 45, -3)(0, 10){5}{\line(0,1){6}}		
		\multiput( 75, -3)(0, 10){5}{\line(0,1){6}}
		\multiput( 30, -3)(0, 5){11}{\line(0,1){3}}		
		\multiput( 90, -3)(0, 5){11}{\line(0,1){3}}		
		\put(  30,-3){\line(0,1){6}} \put(  30,-10){\makebox(0,0){\footnotesize $ \pi$}}
		\put(  60,-3){\line(0,1){6}} \put(  60,-10){\makebox(0,0){\footnotesize $ 2\pi$}}
		\put(  90,-3){\line(0,1){6}} \put(  90,-10){\makebox(0,0){\footnotesize $ 3\pi$}}
		\put( 120,-3){\line(0,1){6}} \put( 120,-10){\makebox(0,0){\footnotesize $ 4\pi$}}		
		\put( 140,-5){\makebox(0,0){\footnotesize $t$}}
		\put(-3,40){\line(1,0){6}} \put(-7,40){\makebox(0,0){\footnotesize $1$}}		
		\put(-10,50){\makebox(0,0){\footnotesize $w(t)$}}		
		\color{green}
		\thinlines    
		\put(-10,0){\line(1,0){40}}
		\qbezier( 30, 0)( 36, 0)(38, 20)
	  \qbezier( 38, 20)(40, 40)(45, 40)
		\put(45,40){\line(1,0){30}}
		\qbezier(75, 40)(80, 40)(82, 20)
	  \qbezier(82, 20)(84, 0)(90, 0)		
		\put(90,0){\line(1,0){45}}
		\color{red}
		\linethickness{1.2mm}
		\put( 45,0){\line(1,0){30}}   \put( 60,9){\makebox(0,0){\footnotesize $I$}}				
\end{picture}
\end{center}
\caption{Illustration of a window function $w\in\C^{\infty}(\RN)$ used in Part II of the proof of Corollary~\ref{cor:WaveEqu} with an interval $I = \big[\frac{3}{2}\pi,\frac{5}{2}\pi\big]$ on which $w$ is identical equal to $1$.}
\label{fig:WindowFunct}
\end{figure}
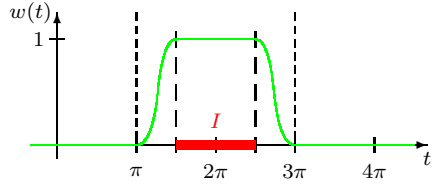

The functions $q(\rho,\cdot)$ and $p_{m}(\rho,\cdot)$ are no longer concentrated on the interval $[\pi,3\pi]$.
Therefore, we multiply $q(\rho,\cdot)$ and $p_{m}(\rho,\cdot)$ with a window function $w\in\C^{\infty}(\RN)$ which is constantly equal to $1$ on an interval $I \subset (\pi,3\pi)$ containing $2\pi$ and which is equal to $0$ outside the interval $[\pi,3\pi]$ (cf. Fig.~\ref{fig:WindowFunct} for an illustration of such a window function).
The obtained functions
\begin{equation*}
	g_{\rho}(t) = w(t)\, q(\rho,t)
	\quad\text{and}\quad
	g_{\rho,m}(t) = w(t)\, p_{m}(\rho,t)	
	\end{equation*}
are equal to zero outside the interval $[\pi,3\pi]$.
In particular, $g_{\rho}\in\C^{\infty}(\RN)$ and satisfies $g'_{\rho}(2\pi) = \frac{\partial q}{\partial t}(2\pi,\rho)\in\RNc$.
Moreover, similarly as in Part~II of the proof of Corollary~\ref{cor:DP_LTI}, the computability of $q$ and properties of the  Poisson kernel imply that $g_{\rho}$ is a computable continuous function.
Indeed by the maximum principle of harmonic functions and \eqref{equ:ApproxByPm}, we have
\begin{align*}
	\left| g_{\rho}(t) - g_{\rho,m}(t) \right|
	= \left| w(t) \right|\, \left| q(\rho,t) - p_{m}(\rho,t) \right|
	&\leq \left| \frac{1}{\pi}\int^{3\pi}_{\pi} \left[ q(\tau) - p_{m}(\tau) \right]\, \frac{\rho}{\rho^{2} + (t-\tau)^{2}}\, \d\tau \right|\\
	&\leq K_{1} \max_{\tau\in [\pi,3\pi]}\left| q(\tau) - p_{m}(\tau) \right|
	\leq K_{1}\, 2^{-N}
\end{align*}
for all $m \geq e(N)$, using that there is a constant $K_{1} > 0$ such that
\begin{equation*}
	\frac{1}{\pi}\int^{3\pi}_{\pi} \frac{\rho}{\rho^{2} + (t-\tau)^{2}}\, \d\tau \leq K_{1}
	\quad\text{for every}\ \rho\in (0,1]\,.
\end{equation*}
So the computable functions $g_{\rho,m}$ effectively converge to $g_{\rho}$ as $m\to\infty$, uniformly for all $t\in [\pi,3\pi]$ and all $\rho\in (0,1]\cap\RNc$.
This verifies that $g_{\rho}\in\CC(\RN)$. 
Overall we thus have $g_{\rho}\in\Mcomp$ for every $\rho\in (0,1]\cap\RNc$.
However, for $\rho = 0$, we have $g_{0}(t) = w(t)\, q(0,t) = w(t)\, q(t)$.
Consequently $g'_{0}(2\pi) = q'(2\pi)$ because $q$ was chosen from $\Mnocomp$, showing that $g_{0} \in \Mnocomp$.

Based on the so-defined function $g_{\rho}$, we define as in \eqref{equ:f_von_g} the functions $f_{\rho}(x) = g_{\rho}\left(\left\|x\right\|\right)$ for all $x\in\RN^{3}$, and Lemma~\ref{lem:McompVcomp} shows that 
\begin{equation}
\label{equ:frhoFaelle}
\begin{array}{lll}
	f_{\rho} \in \Vnocomp & \text{if} & \rho = 0\\
	f_{\rho} \in \Vcomp & \text{for all} & 0 < \rho\leq 1\;.
\end{array}	
\end{equation}

The remainder of the proof follows almost exactly the lines in the proof of Corollary~\ref{cor:DP_LTI}.
We assume that there exists a Turing machine $\TM_{1}(f)$ with input $f \in \Vc$ and which halts if and only if the input $f$ belongs to $\Vnocomp$.
Then we consider the Turing machine $\TM_{>0}(\lambda)$ with input $\lambda\in [0,1]\cap\RNc$ defined as in Part~I of the proof of Corollary~\ref{cor:DP_LTI}.
Similarly as in part one of this theorem, we can now construct a Turing machine $\TMD(\lambda)$ with $\lambda\in [0,1]\cap\RNc$ and which always halts with the two output states
\begin{equation*}
	\TMD(\lambda) = 
	\left\{\begin{array}{lll}
	0 & \text{if} & \TM_{1}(f_{\lambda})\ \text{halts}\\[0.618ex]
	1 & \text{if} & \TM_{>0}(\lambda)\ \text{halts}.
	\end{array}\right.
\end{equation*} 
Nevertheless, such a Turing machine does not exist because the subset $\left\{ \lambda\in [0,1]\cap\RNc : \lambda>0 \right\}$ is known to be not decidable \cite{PourEl_Computability}.
So we obtained a contradiction, proving that the Turing machine $\TM_{1}$ does not exist.
\end{IEEEproof}

At the moment, big technology companies are putting a lot of effort into building virtual representations of the real physical world on digital computers.
With digital representations, they want to solve real world problems in areas like robotics, autonomous driving, autonomous flight systems, and network optimization for communication systems. 
"Metaverse" or "digital twins" are the most discussed approaches in this direction.
These targeted applications have very critical performance parameters because the digital representation on the Turing machine must be a trustworthy image of the real world.
For non-computable physical processes, this is generally not possible, as the results in Sections~\ref{sec:DecProblLTI}-\ref{sec:Conclusion} show.
Even for computable physical processes, a trustworthy digital approximation may be far too complex from a practical point of view \cite{PB_IEEE_TSP21}.

We would like to come back to the two questions by Daniel Kahneman on human-machine interaction and artificial general intelligence (AGI) which were shortly mentioned at the end of Section~\ref{sec:Contribution}.
According to Daniel Kahneman \cite{Fridman_OnlineKahneman}, it would be very interesting to find problems which can not be solved by machines (even on the basis of artificial intelligence), but which can be solved by humans.
Moreover, it would be interesting to find algorithms which are able to autonomously detect such situations, i.e. problems which for some concrete parameters can only be solved by humans, but not by machines.
Kahneman calls this "problematic situations".
The results of Sections~\ref{sec:DetectionUpperBound} and \ref{sec:WaveEquation} answer these questions for all possible machines (including artificial intelligence or AGI machines) which are implemented on digital hardware, i.e. on Turing machines.
Please note that at the moment, all commercially available AI applications are implemented on digital hardware.
Concretely, the results of Sections~\ref{sec:DetectionUpperBound} and \ref{sec:WaveEquation} show that for certain simple, computable, continuously differentiable functions $u$, there exists no Turing machine with input $n\in\NN$ and whose output is a computable number $x_{n} \in \left( \left\|u'\right\|_{\infty} ,  \left\|u'\right\|_{\infty} + 1/2 \right)$.
Nevertheless, a human can certainly solve this problem by creative work.
If, on the other hand, the function $u$ is chosen such that $\left\|u'\right\|_{\infty} \in \RNc$, then there exists a Turing machine which can solve this problem.
In other words, $\left\|u'\right\|_{\infty} \notin \RNc$ is a problematic situation in the sense of Daniel Kahneman, and the results of Section~\ref{sec:DecProblLTI} show that there exists no Turing machine which is able to recognize this problematic situation.
These phenomena can also be shown for solutions of the wave equation (cf. Section~\ref{sec:WaveEquation}).
It is clear that the two questions by Daniel Kahneman address key functionality requirements of future robotic systems in critical environments.
For instance, NASA is developing a vehicle system manager (VSM) for the lunar gateway space station. This robotic system needs to have the corresponding human-robotic collaboration functionality \cite{IEEESpectrum_Lunar}.

It should be emphasized that the previous discussion applies only to machine implementations on digital hardware.
For other hardware platforms, like neuromorphic or quantum computing, the corresponding questions are open.

\section{Discussion and Future Problems}
\label{sec:Conclusion}

It is known that the first derivative of a computable continuously differentiable function is generally not computable.
Using the framework of the Zheng-Weihrauch hierarchy, this paper classified precisely the non-computability of $u'$ and $\left\| u' \right\|_{\infty}$ for functions $u\in\Uc$.
Then it was shown that there exists \emph{no} Turing machine which is able to detect whether, for a given function $u\in\Uc$, it is possible to compute its first derivative at a certain point $\tau\in\TT$ under an effective control of the approximation error.
In other words, there is no way that a digital machine can autonomously detect whether $u'(\tau)$ is a computable number from observing the input $u \in \Uc$.
Even modern machine learning algorithms will never be able to learn to distinguish between functions in $\Uc$ with a computable first derivative and with a non-computable first derivative, respectively.
Even further, it was shown that no Turing machine exists which is able to detect an upper bound for $\left\| u' \right\|_{\infty}$ for all $u\in\Uc$. 
The paper also presented a similar negative result concerning the detectability of non-computability for solutions of the wave equation.

The results of this paper show fundamental limitations of digital computers for simulating continuous physical problems, in particular for simulating analog circuits and systems.
We would like to discuss some ways to circumvent these problems.

1) \emph{Analog computing:} 
Our results are based on the computing model of a Turing machine which is in fact a theoretical model for any algorithm which can be implemented on a digital computer. For analog computers, these restrictions do not apply.
So analog computing might have advantages over digital signal processing for some computational problems (such as the detection problem discussed in this paper), which cannot be solved on digital computers.
Additionally, analog computing is known to have potentially significant advantages in terms of energy consumption, computation time and bandwidth efficiency; active research in this field still exists (see, e.g.,  
\cite{PaulHuper_CAS94,Suh_LowPowerDFT_11,Gu_CAS12,Haensch_DeepLearningAnalog_ProcIEEE19,Udayanga_MaxwellAnalog_CAS19,Bournez_SurveyComputability_2021}).
Moreover, novel approaches and materials have recently stimulated new research on analog computing \cite{Silva_Science14,ZangenehNejad_NatCommun19}.
With that said, we believe that it is a very interesting research area and would like to identify more problems which can, in principle, be solved by means of analog computers, but not with digital hardware, and vice versa.

2) \emph{Restriction of the input set:}
We might restrict the input to a certain subset $\Ud \subset \Uc$.
Then Problem~\ref{prob:GeneralDProblem} might be solvable for this subset.
Results in this direction were already presented in \cite{BP_IEEECAS20}.
There, two subsets $\mathcal{U}_{i} \subset \Uc$, $i = 1,2$, defined by
\begin{align*}
	\mathcal{U}_{1} &= \left\{ v \in \Uc : v'' \in L^{1}_{\mathrm{c}}(\TT) \right\}\\
	\mathcal{U}_{2} &= \left\{ v \in \Uc : v \in H^{s}_{\mathrm{c}}(\TT),\ s>3/2 \right\}
\end{align*}
were given so that for any $u\in\mathcal{U}_{i}$, $i=1,2$, 
its first derivative $u'$ is a computable continuous function. 
So in particular, for any $u\in\mathcal{U}_{i}$, $i=1,2$, one has $u'(0)\in\RNc$.
Therein, the set $H^{s}_{\mathrm{c}}(\TT)$ stands for all functions which are computable in the usual Sobolov space $H^{s}(\TT) = W^{s,1}(\TT)$, and we refer to \cite{BP_IEEECAS20} for a precise definition of the subsets $\mathcal{U}_{1}$ and $\mathcal{U}_{2}$.
In view of these results, we think it is an interesting problem of future research, characterizing subsets of $\Uc$, so that for these subsets Problem~\ref{prob:GeneralDProblem} will have a positive answer.

3) \emph{Dependence on the description of the argument:}
In Section~\ref{sec:DetectionUpperBound}, we investigated the problem of detecting upper bounds on $\left\|u'\right\|_{\infty}$.
There, we considered only possible upper bounds $\lambda(u)$ for functions $\left\|u'\right\|_{\infty}$, $u\in\Uc$, which are computable functions by themselves.
As for any computable function $\lambda$, this implies that its value $\lambda(u)$ does not depend on the actual description of its argument $u$.
However, one could also omit the restriction of $\lambda$ to be computable and consider instead computable mappings from descriptions of the input onto descriptions of the output.
This approach might yield new insights but needs to be left for future research.

4) \emph{Computing models with real numbers:}
Turing machines can only compute with rational numbers.
This is the reason why they can't \emph{exactly} calculate the solution of even very simple differential equations (cf. Sec.~\ref{sec:CircuitsAndComputation}).
However, there do exist computing models which are able to operate with arbitrary real numbers \cite{BSSComputingModel_89}.
These models cannot be implemented with present-day digital technology, in contrast to Turing machines discussed before.
Nevertheless, it is an interesting question whether these computing models allow in principle an exact simulation of arbitrary networks.
Yet, based on the results in \cite{BSSComputingModel_89}, it can easily be shown that this theoretical computing model (a so-called \emph{Blume-Shub-Smale (BBS) machine}) is also too weak to allow for an exact calculation of solutions of very simple differential equations in finitely many computation steps. 
Indeed, it was shown in \cite{BSSComputingModel_89} that every BSS-machine which stops after finitely many iterations, always defines at its output a polynomial or rational function on certain subintervals of its domain.
So the output of a BSS-machine can never be an exponential function.
This shows that, similarly as for Turing machines (cf. Sec.~\ref{sec:CircuitsAndComputation}), solutions of arbitrary RLC-networks cannot be calculated exactly on BSS-machines.
To put it differently, the output $p$ of an BSS-machine always satisfies the differential equation $p^{(n)}(t) = 0$ for all $t \in I$, where $I$ is a sub-interval of the domain of the BSS-machine and $n\in\NN$ is a certain integer.
However, these simple differential equations are basically irrelevant for describing any non-trivial analog circuit.
So even a much more powerful computing model, as the one in \cite{BSSComputingModel_89}, does not allow us to compute solutions of simple differential equations exactly with finitely many iterations.
Thus, even with such powerful computing models one has to rely on approximation procedures which allow for an effective control of the approximation error, i.e. one still has to investigate whether certain operations are computable (with respect to the actual computing model) or not.

5) \emph{General Purpose Analog Computer by Shannon:}
Yet another interesting approach to be investigated in further research is Shannon's famous General Purpose Analog Computer (GPAC) \cite{Shannon_GPAC41}. 
Even though it was recently found that this computing model is equivalent to the Turing machine as far as computability and complexity is concerned \cite{Bournez_SurveyComputability_2021,Garca_AnalogComputers2003}, it would be interesting to investigate the detection problems of this paper within this framework.

6) \emph{Computable sets:}
There might be an other interesting way to look on the problems discussed in this paper, namely by using the notion of \emph{computable sets}  (see, e.g., \cite{BrattkaHertlingWeihrauch_CompAnalysis08}).
Using this approach, one may consider the set
$$A = \left\{ f \in \Uc\ :\ f'\ \text{is computable from}\ f\right\}$$
of all functions in $\Uc$ for which the first derivative is again a computable function.
Then the question whether the computability of $f'$ is decidable can be reformulated as the question whether $A$ is a computable set.
Nevertheless, whereas \cite{BrattkaHertlingWeihrauch_CompAnalysis08} considers computable sets in $\RN^{n}$, our sets belong to infinite dimensional spaces so that the above reformulation of our problem might not be that straight forward and left, therefore, for further research.

Moreover, it might be interesting to investigate how the results of this paper and of \cite{PB_IEEE_TSP21} imply limitations on the ability to build trustworthy digital representations of continuous real world problems based on metaverse and digital twin approaches.

\appendix

\subsection{The Poisson integral}
\label{sec:FirstAppendix}

Let $f \in L^{1}(\TT)$ be a function on $\TT$, then for any $0 \leq r < 1$ and $t\in\TT$, its \emph{Poisson integral} is defined to be 
\begin{equation}
\label{equ:PoissonIntegral}
	f(r,t)
	= \left( \Op{P}_r f\right)(t)
	= \frac{1}{2\pi} \int_{\TT} f(\tau)\, \mathcal{P}_{r}(t-\tau)\, \d\tau\;,
\end{equation}
with the Poisson kernel
\begin{equation*}
	\mathcal{P}_{\rho}(t) = \frac{1-\rho^{2}}{1-2\rho\cos(t) + \rho^{2}}\;.
\end{equation*}
It is well known (cf., e.g., \cite[Theorem~11.8]{Rudin}) that for $f \in \C(\TT)$, $\Op{P}_{r} f$ satisfies the maximum-modulus principle
\begin{equation}
\label{equ:maxModulusPoission}
	\max_{t\in\TT} \left| \left(\Op{P}_{r} f\right)(t)\right| \leq \left\| f \right\|_{\infty}\;,
	\quad\text{for all}\ 0\leq r < 1\;,
\end{equation}
and $\lim_{r\to 1} \left\| \Op{P}_{r} f -f \right\|_{\infty} = 0$.
The later implies in particular that $\lim_{r\to 1} \left\| \Op{P}_{r} f \right\|_{\infty} = \left\| f \right\|_{\infty}$.
If $f$ is even continuously differentiable on $\TT$, i.e. $f \in \C^{1}(\TT)$, one additionally has $\lim_{r\to 1} \left\| \Op{P}_{r} f' \right\|_{\infty} = \left\| f' \right\|_{\infty}$
and 
\begin{equation}
\label{equ:PoissonDerivative}
	\left( \Op{P}_{r} f' \right)(t) = \frac{\d}{\d t} \left(\Op{P}_{r} f\right)(t)\;,
	\quad t\in\TT\;.
\end{equation}

\begin{lemma}
\label{lem:AppxPoission}
Let $f \in \C^{1}(\TT)$ and let $\left\{ f_{n} \right\}_{n\in\NN}$ be a sequence in $\C^{1}(\TT)$ so that $\lim_{n\to\infty} \left\| f - f_{n} \right\|_{\infty} = 0$.
If there exists a $g \in \C(\TT)$ so that $\lim_{n\to\infty} \left\|g - f'_{n} \right\|_{\infty} = 0$,
then $f' = g$. 
\end{lemma}

\begin{IEEEproof}
Since $\Op{P}_{r} : \C(\TT) \to \C(\TT)$ is bounded (cf. \eqref{equ:maxModulusPoission}) for every $0 \leq r < 1$, the assumptions of the lemma imply
\begin{equation*}
	\left( \Op{P}_{r} g \right)(t)
	= \lim_{n\to\infty} \left( \Op{P}_{r} f'_{n} \right)(t)
	= \lim_{n\to\infty} \tfrac{\d}{\d t} \left( \Op{P}_{r} f_{n} \right)(t)\,,
	\quad t \in\TT\;,
\end{equation*} 
and
\begin{equation*}
	\lim_{n\to\infty} \left( \Op{P}_{r} f_{n} \right)(t) = \left( \Op{P}_{r} f \right)(t)\,,
	\quad t \in\TT\;.
\end{equation*}
By the properties of the Poisson integral, this also holds for the derivative, i.e.
\begin{equation*}
	\lim_{n\to\infty} \tfrac{\d}{\d t}\left( \Op{P}_{r} f_{n} \right)(t)
	= \tfrac{\d}{\d t} \left( \Op{P}_{r} f \right)(t)
	= \left( \Op{P}_{r} f' \right)(t)\,,
	\quad t \in\TT\;,
\end{equation*}
where the last equality follows from \eqref{equ:PoissonDerivative} because $f \in \C^{1}(\TT)$.
Consequently, we have $\left( \Op{P}_{r} g \right)(t) =  \left( \Op{P}_{r} f' \right)(t)$ for all $0 \leq r < 1$ and $t \in \TT$.
Since $\Op{P}_{r}$ is linear, we therefore have
\begin{equation*}
	0 = \lim_{r\to 1} \left[\Op{P}_{r}(f' - g)\right](t) = f'(t) - g(t)\;,
	\quad \text{for all}\ t\in\TT\;,
\end{equation*}
proving that $f' = g$.
\end{IEEEproof}

\section*{Acknowledgment}

H.~Boche would like to thank Ignacio Cirac for hints and questions regarding the computability of physical theories, in particular, of physical parameters.
He further thanks Eike Kiltz and Alexander May for discussions and questions about quantum algorithms for the computation of hidden subgroups of computable functions and related problems in cryptanalysis of “post-quantum cryptography”, and he thanks Frank Fitzek for interesting discussions on metaverse and digital twin approaches.

\end{document}